\documentclass[a4paper,fleqn]{cas-dc}

\usepackage[numbers]{natbib}
\usepackage{algorithm}
\usepackage{algorithmic}
\usepackage{tabularx}
\usepackage{capt-of}

\usepackage{soul}
\usepackage{xcolor}
\sethlcolor{yellow!30}

\def\tsc#1{\csdef{#1}{\textsc{\lowercase{#1}}\xspace}}
\tsc{WGM}
\tsc{QE}
\tsc{EP}
\tsc{PMS}
\tsc{BEC}
\tsc{DE}

\begin{document}
\let\WriteBookmarks\relax
\shorttitle{Remote Attestation of Smart Meters}
\shortauthors{A. Shakiba et~al.}

\title [mode = title]{Resource-constrained Remote Attestation of Smart Meters Using Merkle Trees in a Trusted Execution Environment}                      



\author[1]{Amirhossein Shakiba}[type=editor,
                        auid=000,bioid=1,
                        orcid=0009-0003-4941-7165]
\ead{amirhosein.shakiba@modares.ac.ir}

\author[2]{Hamid Reza Ghaeini}[
auid=000,bioid=2,
orcid=0000-0003-2614-7900]
\cormark[1]
\ead{ghaeini@ut.ac.ir}
\ead[url]{https://profile.ut.ac.ir/~ghaeini}

\author[1]{Sadegh {Dorri Nogoorani}}[orcid=0000-0001-7406-112X]
\cormark[1]
\ead{dorri@modares.ac.ir}


\affiliation[1]{organization={Faculty of Electrical and Computer Engineering, Tarbiat Modares University},
                city={Tehran},
                country={Iran}}

\affiliation[2]{organization={Fouman Faculty of Engineering, College of Engineering, University of Tehran},
                city={Tehran},
                country={Iran}}

\cortext[cor1]{Corresponding author}


\begin{abstract}
Smart meters are pivotal to the functioning of smart grids, as they measure and transmit energy consumption data that includes sensitive information about household energy usage patterns. While this data offers significant financial benefits, it can also reveal detailed insights into residents' daily routines, making the security of smart meters a critical concern. This paper introduces a novel remote attestation method for smart meters that ensures the integrity and privacy of data exchanged between smart meters and central servers while detecting unauthorized tampering. Our approach structures the communication data into a Merkle Tree—a hash data structure that facilitates efficient data integrity and authenticity verification. All operations involved in constructing the Merkle root and hashing critical data are executed in the secure environment of a Trusted Execution Environment (TEE) using OP-TEE, thereby maintaining the integrity and security of the smart meter's data. Comparing our experimental results with the related works highlights the superiority of our method in terms of resource efficiency, security robustness, and the feasibility of implementation on smart meters and resource-constrained devices, thereby enhancing the overall security and reliability of smart grid systems.
\end{abstract}




\maketitle

\section{Introduction}
\label{sec:introduction}

Smart meters play a crucial role in the functionality of a smart grid; without them, the seamless flow of the smart grid chain is disrupted. In addition to their primary function of calculating and measuring electricity consumption, smart meters serve as multifunctional devices that also collect and store important data, including time-of-use information, voltage levels, power quality metrics, fault diagnostics, and billing details~\cite{kumar2018lightweight}.

\begin{figure*}[!t]
	\centering
	\includegraphics[width=0.8\textwidth]{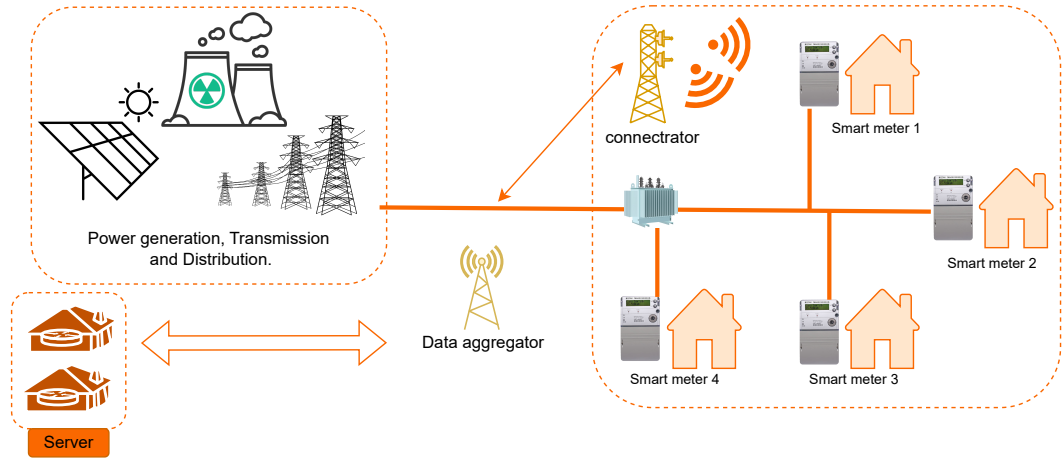}
	\caption{Smart grid architecture and its components.}
	\label{smrtgrid}
\end{figure*}

The data collected by smart meters is important and sensitive and is essential to users' privacy. Access to this information by unauthorized people carries the risk of misuse and other potential dangers. For example, this data can reveal when people are away from home or asleep, which underlines the importance of protecting and managing sensitive information responsibly.

Smart meters can be connected to the smart grid in various ways, e.g., via a wired network of power lines or wireless communication technologies.
Each of these methods is subject to certain communication standards that enable interaction with servers. Given the central role of smart meters in the smart grid, ensuring the security of these meters has become an integral aspect of the entire grid~\cite{kabalci2016survey}.
Various methods are used to protect the integrity of smart meters, including encryption, authentication, access control, intrusion detection, prevention systems, and data anonymization.
In addition, data segmentation within smart grid components, including smart meters, is used to enhance overall security measures~\cite{karopoulos2017towards}.

The work in~\cite{11028800} takes a fairly broad view of smart meter security, including their structure and possible weak points: cyber, physical, and cases that combine the two. Vulnerabilities can also be categorized into being at the hardware level, in software, or across the communication network. According to~\cite{11028800}, several practical measures should be considered for secure deployment, among them are strong authentication, integrity checks on metering data, confidentiality and privacy safeguards, and compliance with existing regulations. It should be stressed that at none of these layers is sufficient on its own. Smart meters sit at the boundary between digital and physical infrastructure, and security has to be handled across both infrastructures at the same time.

Our proposed method leverages ARM TrustZone's Trusted Execution Environment(TEE) to establish a secure and isolated workspace for executing sensitive operations, thereby significantly enhancing the security of smart meters. Central to our approach is a Merkle Tree, utilized as the fundamental data structure for securely transmitting original smart meter data. This design ensures immediate detection of any data modifications, maintaining the integrity and authenticity of transmitted information. Furthermore, the integration of Merkle Tree root with a Nonce and HMAC hashing provides a robust mechanism for verifying the remote attestation system's trustworthiness, ensuring the data's integrity. To demonstrate the practical benefits of our method, we conducted a comparative evaluation against traditional encryption-based and signature-based verification techniques. Encryption-based methods, while effective in securing data transmission, introduce significant computational overhead and latency, making them unsuitable for resource-constrained environments like smart meters.

Similarly, signature-based schemes offer robust data integrity and authenticity assurances but suffer from high computational costs, particularly with large data volumes or frequent attestation requests. In contrast, our Merkle Tree-based authentication within a TEE enables rapid and scalable data integrity verification without the heavy cryptographic computations required by traditional methods. Merkle Trees offer compact and verifiable proofs of authenticity, significantly reducing processing time and memory usage. Additionally, executing critical operations within a TEE ensures a secure and isolated environment, minimizing risks of tampering and unauthorized access. Our experimental results confirm that Merkle Tree construction and verification processes are scalable as the tree size increases, whereas encryption and signature-based methods experience significant performance declines under similar conditions. Consequently, our method not only enhances security and reliability but also ensures operational efficiency, positioning it as a superior choice for modern smart grid systems. Altogether, these innovations establish a comprehensive and resilient framework for the secure management of smart meter data.

The rest of the paper is organized as follows. First, in section~\ref{sec:background}, we provide a background on the topics presented within the paper. We summarize related work in section~\ref{sec:relatedworks}. We describe our proposed method in Section~\ref{sec:proposedmethod}.  We present our evaluation results in section~\ref{sec:evaluation}.  Finally, we conclude in section~\ref{sec:conclusion}.

\section{Background}
\label{sec:background}
This section provides background information on key topics discussed in the paper, including the smart grid, types of smart meters, trusted execution environments, remote attestation, and Merkle Trees, to provide the required background knowledge.

\subsection{Smart Grid}
The smart grid promotes bidirectional communication between electricity suppliers and their customers and enables the optimal integration of new energy solutions, such as solar and wind energy, into the existing infrastructure~\cite{khalaf2024survey}.
By integrating smart household meters into the smart grid, households can manage their electricity consumption efficiently~\cite{kohout2023smart}.

Smart grids optimize energy distribution and management through components like the dispatching center, which oversees generation and distribution while analyzing smart meter data. Connectors and data aggregators facilitate seamless communication and integration across the grid, enabling efficient operations, especially in smart homes. While smart meters enhance real-time monitoring and management, they raise security concerns, highlighting the need for secure communication. Together, these elements ensure efficient and reliable energy management in modern smart grids.

A major difficulty with smart grids is the complexity of utilizing different protocols and functions for the connected end devices, including smart meters and concentrators~\cite{hseiki2024secure}.
This complexity leads to vulnerabilities that require extensive source code to implement complicated systems. In addition, devices connected to home networks are inherently more vulnerable to attack than devices connected directly to controlled networks~\cite{tatipatri2024comprehensive}.

The main issue with smart meters is their vulnerability to cyber-attacks, which can lead to unauthorized access, power disruptions, and potential network damage. On the other hand, ensuring software integrity protects the firmware from unauthorized changes, while data integrity ensures the accuracy and consistency of transmitted information. Robust encryption, regular updates, and secure communications are critical to mitigate these risks, especially as smart meters are increasingly integrated into the broader Internet of Things (IoT) ecosystem~\cite{paverd2012hardware}.

In the broader landscape of smart management systems, modern distribution grids employ a range of metering and monitoring devices that can be categorized into different classes of smart management devices.
These devices are substantially changing the grid landscape, with four main categories identified: smart energy sensors, smart feed-in meters, smart distribution transformer monitoring devices, and remote control switches. This discussion will focus primarily on the importance of smart energy sensors and smart meters, highlighting their critical role in the evolving energy management ecosystem~\cite{chakraborty2021smart}.

\subsection{Smart meter}
Electricity metering is critical to the operation of electrical systems and plays an important role in energy management, energy billing, and ensuring the reliability and safety of the system~\cite{mirshekali2021real}.
With the development of smart grids and the growing need to improve electricity management and maximize energy efficiency, the deployment of smart meters has become increasingly important~\cite{wang2018review}. 

Smart meters, an evolution of conventional energy meters, have new features designed to provide innovative solutions, particularly in the collection of relevant information on electricity consumption and two-way communication, allowing them to interact dynamically with centralized systems~\cite{avancini2021new}.
These features facilitate remote disconnection and reconnection of electrical loads, as well as the monitoring and controlling of user devices in the context of smart buildings~\cite{garcia2017power}.

The integration of various sensors and control devices, supported by particular communication infrastructures, underlines the versatile role of smart meters in efficiently managing energy demand and loads~\cite{zhang2020privacy}.
In addition, smart meters play a central role in supporting dynamic pricing mechanisms and facilitating the bi-directional flow of electricity between residential units (or microgrids) and the general electricity grid.
However, the widespread deployment of smart meters raises significant privacy concerns as they can collect comprehensive data on household activities~\cite{rincon2020securing}.

While smart meters have numerous potential benefits for consumers and utilities, there are also significant concerns about security and privacy risks.
Primary concerns include unauthorized access to data, remote control vulnerabilities, and inherent security and privacy flaws~\cite{anderson2011smart, avancini2021new}.
In order to mitigate these risks, robust security measures need to be implemented, such as strong encryption, strict access controls, increased public awareness, and the creation of a strong legal framework. 
Addressing these concerns is crucial for broad acceptance and trust in smart meter technology. Ultimately, the decision to roll out smart meters is influenced by assessing the perceived benefits and potential risks~\cite{mahmud2015survey, mirshekali2021real}.

\subsection{Trusted Execution Environment}
The Trusted Execution Environment (TEE) is an evolving technology that improves the security of computer systems by creating a secure and isolated environment for the execution of sensitive applications. 
TEE is often referred to as a \emph{secure environment} or \emph{trust zone} and protects the code and data of active applications from unauthorized access, tampering, and intrusion.

The rapid growth of the Internet of Things requires intelligent systems that often run on ARM-based processors~\cite{guan2017trustshadow}.
To improve software security, ARM has developed TrustZone technology, which creates a foundation of trust by dividing resources into secure and insecure parts. 
Despite the presence of untrusted third parties, TrustZone effectively secures high-quality data and enables a reliable execution environment~\cite{schonstedt2022safetee}.

ARM Cortex-M23 and Cortex-M33 based microcontrollers have emerged as the dominant choice for resource-constrained IoT deployments, combining hardware-enforced TrustZone-M with ultra-low power consumption and a competitive price; devices such as the Microchip SAM L11, Nuvoton NuMicro M2351, and NXP LPC55S6x illustrate this trend across different vendors, each delivering secure boot, isolated execution, and real-time performance on a single low-cost chip. Smart meters represent one of the most practical and cost-sensitive deployments of this technology, where the economics of scale make every per-unit decision consequential: the STMicroelectronics STM32L5, a Cortex-M33 running at 110 MHz with 512 KB Flash and 256 KB SRAM, exemplifies how TrustZone-M can be put to work in exactly this context, integrating the TEE directly on-chip to eliminate the need for a discrete Secure Element or HSM while drawing remarkably little power and process---a combination well-suited to battery-assisted or energy-harvesting meter designs. Within that isolated environment, the meter's most sensitive operations—energy measurement, billing computation, cryptographic key management, and secure over-the-air firmware updates—execute entirely apart from the untrusted application world, preserving both integrity and confidentiality without dedicated security hardware, and across a national rollout of millions of units, even a modest reduction in per-unit bill-of-materials cost can translate to substantial aggregate savings, making this class of ARM-based TEE architecture not only technically adequate but economically compelling for large-scale infrastructures.

\subsection{Remote attestation}
Remote attestation is a cybersecurity process that can be used to verify the trustworthiness of systems or applications on other devices by digitally verifying identity and integrity. 
It is crucial for confidential data processing and ensures that the devices comply with the expected settings. Key components include the verification server (verifier), the verification client (credential sender), the key broker service (KBS), which stores secrets and releases them after verification, and the key broker client (receives keys from the KBS). Remote attestation enables dynamic responses to vulnerabilities and more comprehensive policy authentication compared to Trusted Platform Modules (TPM), which are essential for secure data processing.

\subsection{Merkle Tree}
Merkle Trees are cryptographic data structures that enhance the efficiency and security of data validation and storage. They use a root hash derived from individual data hashes, enabling fast, bandwidth-efficient validation of specific data within large datasets without accessing the entire set~\cite{arasu2021fastver}. 

This structure reduces storage requirements, supports parallel verification, and is ideal for secure, distributed systems requiring efficient integrity checks and tamper detection. Merkle Trees excel in secure outsourced data systems by enabling effective updates and verification. They construct a tree of cryptographic hashes, combining child node hashes into a single root hash, which is signed by the data owner. Each update generates a new root hash that ensures data integrity, making Merkle Trees well-suited for applications like secure key-value outsourcing~\cite{el2023towards}.

\section{Related Works}
\label{sec:relatedworks}

This section provides an overview of prior research on smart grid technologies, with a particular focus on smart meters and their associated security challenges. The analysis critically examines existing methods for securing smart meters and contextualizes them against the novel approach proposed in this paper.

A range of methodologies has been explored to enhance the security of smart meters, with special attention to safeguarding user data against potential breaches. Central to these efforts are mechanisms that address core security principles, including authentication, authorization, data protection, privacy, and regulatory compliance. Through these mechanisms, researchers aim to mitigate vulnerabilities that could compromise the integrity of smart grid systems. Furthermore, our review highlights how utility companies can leverage robust security frameworks to protect sensitive information, ensure system reliability, and foster consumer trust. Emphasis is placed on the comparative strengths and limitations of existing techniques, laying the groundwork for our proposed.

\subsection{Authentication and Authorization}
The integration of compression and authentication in advanced metering infrastructure (AMI) addresses key challenges in managing smart meter data efficiently and securely~\cite{lee2019unified}. Prior research has explored separate domains of compression and authentication. Compression techniques, such as wavelet-based methods, SVD, and compressive sensing (CS), reduce data volume and processing demands, while authentication mechanisms like cryptographic methods and physical-layer approaches ensure data integrity. However, cryptographic methods are computationally intensive, and lightweight physical-layer techniques may fall short against closely matched intrusion signals.
The authors propose a unified CS-based framework for simultaneous compression and authentication. It employs a shared measurement matrix as a secret key and residual error-based hypothesis testing, achieving high energy efficiency, reduced computational complexity, and low authentication error rates under practical conditions. The approach is particularly suited for low-cost smart meters. Nevertheless, challenges include potential fidelity loss due to lossy compression, limited adaptability to diverse attack scenarios, and complexity in real-world deployment~\cite{lee2019unified}.

The implementation of smart grids and AMI has driven significant research into secure communication frameworks, particularly in key management and authentication systems. Among these, the logical key hierarchy (LKH) mechanism~\cite{abdalzaher2023survey} stands out as a robust approach for scalable and efficient key management in dynamic SG environments. LKH employs a tree-based structure where each user maintains a copy of the keys for their leaf and all nodes leading to the root. This hierarchical arrangement enables secure key distribution for large-scale demand response projects, significantly reducing storage and communication costs. Furthermore, the method allows multiple projects to share new key sets, enhancing scalability and adaptability in rapidly changing environments. LKH ensures that the rekeying process for users joining or leaving a group does not affect the overall performance of other nodes by optimizing rekeying operations through a multigroup key graph structure. However, the mechanism is not without limitations. In cases of node compromise, the multicast group rekeying process can become resource-intensive, requiring a delicate balance between the number of transmissions and storage.

Another noteworthy advancement is the integration of physically unclonable functions (PUFs) in hardware-oriented authentication systems~\cite{al2023physical}. PUFs exploit the intrinsic randomness in hardware manufacturing to generate unique cryptographic keys, providing a robust defense against replication and impersonation attacks. Recent research has extended this concept by combining PUFs with identity-based authentication, creating a hybrid framework that leverages both symmetric cryptography and identity-based systems. For example, PUF-based frameworks eliminate the need for additional volatile memory systems, as secret keys are derived from hardware properties instead of being stored. This approach has proven effective in mitigating denial-of-service (DoS) attacks and reducing communication latency by optimizing key management at the hardware level. Additionally, the integration of PUFs with Hamming code has demonstrated reduced overhead in data transmission while maintaining strong security. However, challenges persist in terms of real-time computational demands and the complexity of integrating PUFs into existing AMI systems. The cost and technical requirements for deploying PUF-based systems, particularly in low-cost smart meters, remain areas for further exploration.

Logical Key Hierarchy (LKH)~\cite{abdalzaher2023survey} and PUF-based systems~\cite{al2023physical} demonstrate the strengths and weaknesses of contemporary advancements in secure communication for SGs. LKH offers better scalability, efficient key distribution, and reduced storage requirements, making it suitable for dynamic environments with diverse user demands. However, it faces challenges in handling resource-intensive rekeying processes during node compromise. On the other hand, PUF-based systems ensure enhanced security, robust resistance to impersonation attacks, and reduced data transmission overhead but face integration challenges due to high costs and computational demands. These approaches represent critical strides toward efficient and secure smart grid communication.

\begin{table*}[!t]
	\centering
	\caption{Comparison of related works on smart meter security.}
    \label{tab:relatedComparison}
    \setlength{\tabcolsep}{3pt}
	\begin{tabularx}{\textwidth}{|p{2.2cm}|p{0.7cm}|X|X|X|}
		\hline
		\textbf{Work} & \textbf{Year} & \textbf{Contribution} & \textbf{Advantages} & \textbf{Disadvantages} \\ \hline
		Unified CS-based Framework~\cite{lee2019unified} & 2019 & Simultaneous compression and authentication using compressive sensing (CS) with residual error-based hypothesis testing. & High energy efficiency, reduced computational complexity, and low authentication error rates. & Potential fidelity loss, limited adaptability to diverse attacks, complexity in deployment. \\ \hline
		Logical Key Hierarchy (LKH)~\cite{abdalzaher2023survey} & 2023 & Scalable and efficient key management using a hierarchical structure for dynamic SG environments. & Reduces storage and communication costs and enhances scalability for large networks. & Resource-intensive rekeying during node compromise, multicast group complexity. \\ \hline
		Physically Unclonable Functions (PUFs)~\cite{al2023physical} & 2023 & Hardware-oriented authentication leveraging unique cryptographic keys from hardware randomness. & Strong resistance to impersonation attacks, reduced data transmission overhead, low latency. & High cost, real-time computational demands, challenges integrating into existing systems. \\ \hline
		Cloud-Edge-End Architecture~\cite{jiang2020lightweight} & 2020 & Localized edge computing for smart meter data processing and fully homomorphic encryption (FHE) for secure data aggregation. & Low latency, high bandwidth utilization, secure encrypted data operations. & Computational overhead, deployment complexity for low-cost meters. \\ \hline
		EPDA Scheme~\cite{alsharif2019epda} & 2019 & Privacy-preserving data aggregation and re-encryption for multi-recipient AMI networks. & Resists collusion attacks, ensures fine-grained access control, and low computational overhead. & Infrastructure integration challenges, scalability issues in large-scale deployments. \\ \hline
		P2P Energy Theft Detection~\cite{salinas2013privacy} & 2013 & Privacy-preserving theft detection using LU and QR decomposition in a P2P network. & Robust privacy protection, scalability in detecting theft, and collaborative computations. & Numerical instability in large networks, limitations in LU decomposition accuracy. \\ \hline
		State Estimation with PPBE~\cite{salinas2015privacy} & 2015 & Privacy-preserving bias estimation for energy theft detection using Kalman filters. & Strong privacy guarantees, robust detection, and faster convergence with decentralized computation. & Computational complexity, reliance on accurate system parameters. \\ \hline
	\end{tabularx}
	
\end{table*}

\subsection{Data Integrity and Confidentiality}
The \emph{cloud-edge-end} architecture~\cite{jiang2020lightweight} is integrated edge computing for localized data processing, reducing the reliance on centralized master stations and improving system responsiveness. This architecture uses IoT proxy terminals at the edge layer to dynamically manage smart meters through a plug-and-play model, enabling low-latency, scalable, and secure device integration. However, potential security risks arise if edge nodes are compromised, and deployment into existing AMI systems poses infrastructure challenges. Another contribution is the use of fully homomorphic encryption (FHE) combined with a lightweight packet reorganization protocol, enhancing data security during aggregation and reducing packet loss through hash-based mechanisms and sliding address windows. In this regard, the \emph{cloud-edge-end} architecture~\cite{jiang2020lightweight} achieves high bandwidth utilization and low latency while allowing secure mathematical operations on encrypted data. Despite its benefits, the computational overhead of FHE and the complexity of large-scale deployment present challenges for low-cost smart meters, underscoring the need for further optimization in real-world implementations.

Another important contribution to the domain of privacy and efficiency in AMI is the Efficient and Privacy-Preserving Data Collection and Access Control (EPDA)~\cite{alsharif2019epda} scheme. It is designed for multi-recipient AMI networks, addressing the limitations of single-recipient schemes. This framework employs a novel proxy re-encryption mechanism to facilitate data aggregation prior to re-encryption. It ensures that aggregated power consumption data (APCD) is accessible only to intended recipients while maintaining fine-grained access control. EPDA achieves low computational and communication overhead by combining lightweight encryption and re-encryption techniques. This approach also prevents disclosing customers’ distribution to unauthorized entities, ensuring fair electricity trading. Moreover, EPDA resists collusion attacks between malicious entities, such as distribution network operators (DNOs) and the data communication company (DCC). However, integrating EPDA into existing systems may require substantial infrastructure changes and complex key management mechanisms.

\subsection{Privacy Considerations}
Energy theft remains a significant challenge in smart grids, causing substantial economic losses. A notable approach is employing peer-to-peer (P2P) computing for privacy-preserving energy theft detection~\cite{salinas2013privacy}. This method leverages distributed algorithms solving linear systems, such as LU and QR decompositions, to identify fraudulent users without requiring users to disclose their private energy consumption data. The P2P framework ensures user privacy by allowing smart meters to compute \emph{honesty coefficients} collaboratively, preserving individual data integrity. It offers robust privacy protection and scalable theft detection across networks of various sizes. However, challenges include numerical instability in large-scale networks with LU decomposition.

The state estimation-based energy theft detection framework~\cite{salinas2015privacy} employs Kalman filters to identify anomalies in microgrid systems. The privacy-preserving bias estimation algorithm (PPBE) extends this by decoupling bias estimation from system state estimation, preserving users’ privacy while efficiently identifying theft. The method separates the computation into two parallel filters: one estimates unbiased system states while the other computes biases using residuals. This decomposition ensures faster convergence and numerical stability. The advantages of this approach include robust detection capabilities, scalability in large microgrid networks, and strong privacy guarantees through decentralized computation. However, the computational complexity of the filters and reliance on accurate system parameters pose challenges for real-world deployment.

In conclusion, our review of existing smart meter security techniques underscores the multifaceted challenges and advancements within the field. Various approaches, including unified compressive sensing frameworks~\cite{lee2019unified}, Logical Key Hierarchy (LKH)~\cite{abdalzaher2023survey}, Physically Unclonable Functions (PUFs)~\cite{al2023physical}, cloud-edge-end architectures~\cite{jiang2020lightweight}, and privacy-preserving data aggregation schemes like EPDA~\cite{alsharif2019epda}, demonstrate significant strides in enhancing authentication, authorization, data integrity, and privacy in smart grids. These solutions offer benefits such as improved scalability, attack resistance, and efficient data management, yet they also present limitations, including computational overhead, integration complexities, and vulnerability to specific attack vectors. Additionally, innovative techniques in energy theft detection, leveraging peer-to-peer computing~\cite{salinas2013privacy} and state estimation-based frameworks~\cite{salinas2015privacy}, highlight the ongoing efforts to balance privacy with effective anomaly detection. As illustrated in Table~\ref{tab:relatedComparison}, each method presents a unique set of advantages and challenges, providing a clear comparison that informs the development of more robust security solutions. This critical analysis highlights the strengths and weaknesses of current technologies, providing a foundation for our proposed method to address existing gaps and enhance the security and reliability of smart grid systems.

\section{Proposed Method}
\label{sec:proposedmethod}

\begin{figure*}[!t]
	\centering
	\includegraphics[width=0.8\textwidth]{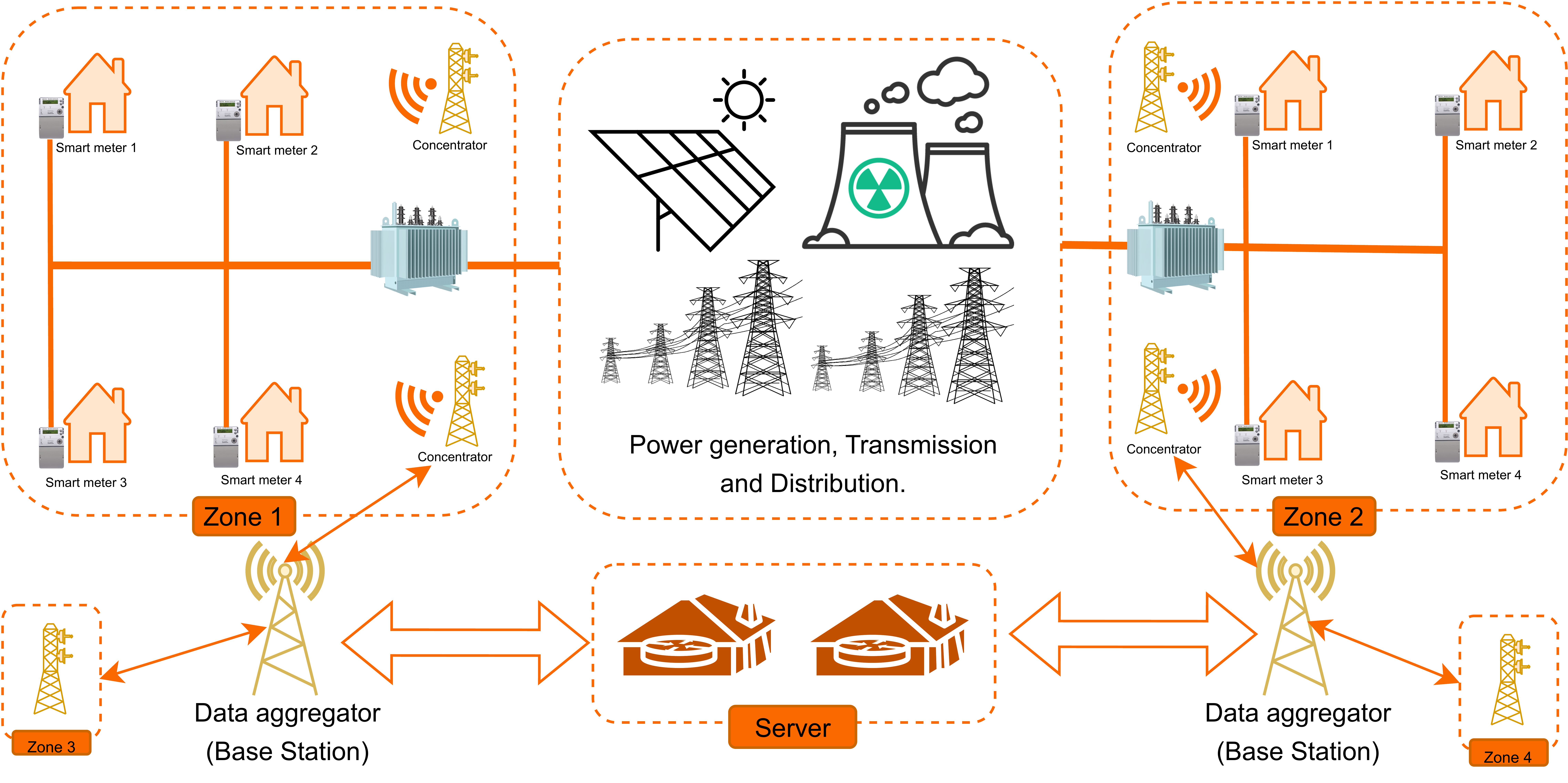}
	\caption{System overview and communications between servers and smart meters.}
	\label{largScale}
\end{figure*}

In this section, we present our work to improve the security of smart meters in smart homes. Figure~\ref{largScale} illustrates an overview of the communication between servers and smart meters. Before a smart meter can be integrated into the grid or verify its identity to the server before transmitting critical data, it must go through a series of steps. In this process, the smart meter's initial and sensitive data, which are static and serve as the leaves of a Merkle Tree, are used to generate the root of the tree. This root is then combined with a random Nonce and hashed as part of the remote attestation process. The resulting data is transmitted to the server, which compares the received data to its stored data for validation. If the remote attestation is successful, the smart meter is authorized to transmit data. If it fails, the smart meter is blacklisted by the server, and its data transmission is blocked.

During this process, both predictable and unpredictable events can occur, such as power outages or loss of communication between meters or with the server. In scenarios where smart meters are responsible for transmitting electricity usage data, the system ensures that any data collected during power or communication outages is securely transmitted to the appropriate systems and servers once connectivity is restored. In order to support this functionality, the smart meter architecture is equipped with a backup battery and temporary memory to retain data during interruptions, ensuring data integrity and continuous operation within the smart grid. 

This approach enhances data transmission reliability and security within the smart grid while ensuring system resilience against potential disruptions, making it a robust solution for large-scale smart meter deployments.

With the proliferation of smart homes, smart meters have proven to be critical components that generate and process important data, making them potential targets for cyberattacks. To mitigate this risk, we have integrated advanced security features into the smart meter, utilizing secure hashing to protect data integrity and significantly delay unauthorized access attempts.

Our proposed method involves constructing a Merkle Tree to organize and securely validate the original raw smart meter data. These components, essential for the accurate and safe operation of the smart meter, form the leaves of the Merkle Tree. Additionally, a symmetric key is embedded on the smart meter's board during manufacturing, ensuring that only the smart meter and the server possess this key. This symmetric key is integrated into the hash operations between the meter and the server, enhancing the overall security by enabling secure and authenticated communication channels.

The root of the tree is transferred to a remote server after hashing for verification to ensure the integrity and authenticity of the smart meter data before it is integrated into the smart grid. All steps, including the formation of Merkle Tree and initial data management, are executed securely within a Trusted Execution Environment (TEE), providing a robust data protection framework.

\subsection{The Process}
In the initial setup of a smart grid, after the installation and placement of the components, Merkle Tree construction phase takes place before the smart meters are activated and connected to the grid. This phase, implemented as a part of the grid's security mechanism, aims to organize and protect the sensitive data of the smart meters. The structure of Merkle Tree plays a crucial role in ensuring data security by preventing unauthorized access to critical information.

The primary data of a smart meter includes the device's unique identifier, initial security information, basic configurations, manufacturer information, the installed software version, the initial energy consumption profile, environmental data, communication parameters, and the history of activities and initial settings. This data, which is essential for the correct and safe operation of the meter before it is connected to the smart grid, forms the leaves of the Merkle Tree.

After the data has been aggregated and hashed and the tree and its root have been formed, a remote attestation procedure is performed to ensure the integrity and security of the data transmitted between the meter and the server. The remote attestation is performed by first calculating the root of Merkle Tree, which contains the meter's sensitive data, and then concatenating it with a Nonce. This concatenation results in a unique and unpredictable value, further processed using the HMAC algorithm and the initially shared symmetric key to generate a secure and unique hash value. This hash value is then sent to the server.

The server performs the same operations with the corresponding data and compares the received hash with the hash stored in its system. The match between the two hashes means the meter and its data are intact and valid, and the remote attestation has been successfully completed so that the meter can join the smart grid. However, the mismatch between the received and calculated hashes indicates a possible problem in the meter firmware or manipulation of the data, resulting in the failure of the remote attestation. In this case, the meter must not be connected to the network, and further investigation is required to identify and rectify the problem or detect possible threats. If the remote attestation fails, the process is repeated to check whether there is a temporary problem. If the error persists, the meter will be blacklisted for further investigation.

Remote attestation is not only used to initially verify smart meters, but it also plays a critical role when updating smart meters and before transmitting critical meter data. Remote attestation is performed before an update begins. This ensures that the meter operates in a factory setting and remains valid and free of unauthorized modifications. Once the attestation is successfully completed, the updated data is sent from the server to the smart meter. Upon completion of the update process, another remote attestation is performed to confirm that the updated meter has been properly integrated into the system and that secure communications have been established. These steps ensure that the meter will continue to function in a secure manner within the smart grid after the update. The smart meter also uses remote attestation to verify the integrity of the outgoing information before transmitting sensitive data. Only after successful attestation is the data transferred to the server. This is a critical step in ensuring that only valid and secure information is transferred to the server, thus maintaining the security of the entire data transfer process.

\subsection{Proposed Architecture}

Our proposed smart meter architecture is based on a standard smart meter with an ARM processor with TrustZone support, as depicted in Figure~\ref{smrtmeter}.
At the hint of this meter is the trusted execution environment or the main processor of the device. It provides secure code execution and protects sensitive data from unauthorized access. 
This execution environment is characterized by a high level of security, a reliable computational base, isolation, and secure execution. It includes the following components:

\begin{center}
	\includegraphics[width=0.95\columnwidth,height=0.7\textheight,keepaspectratio]{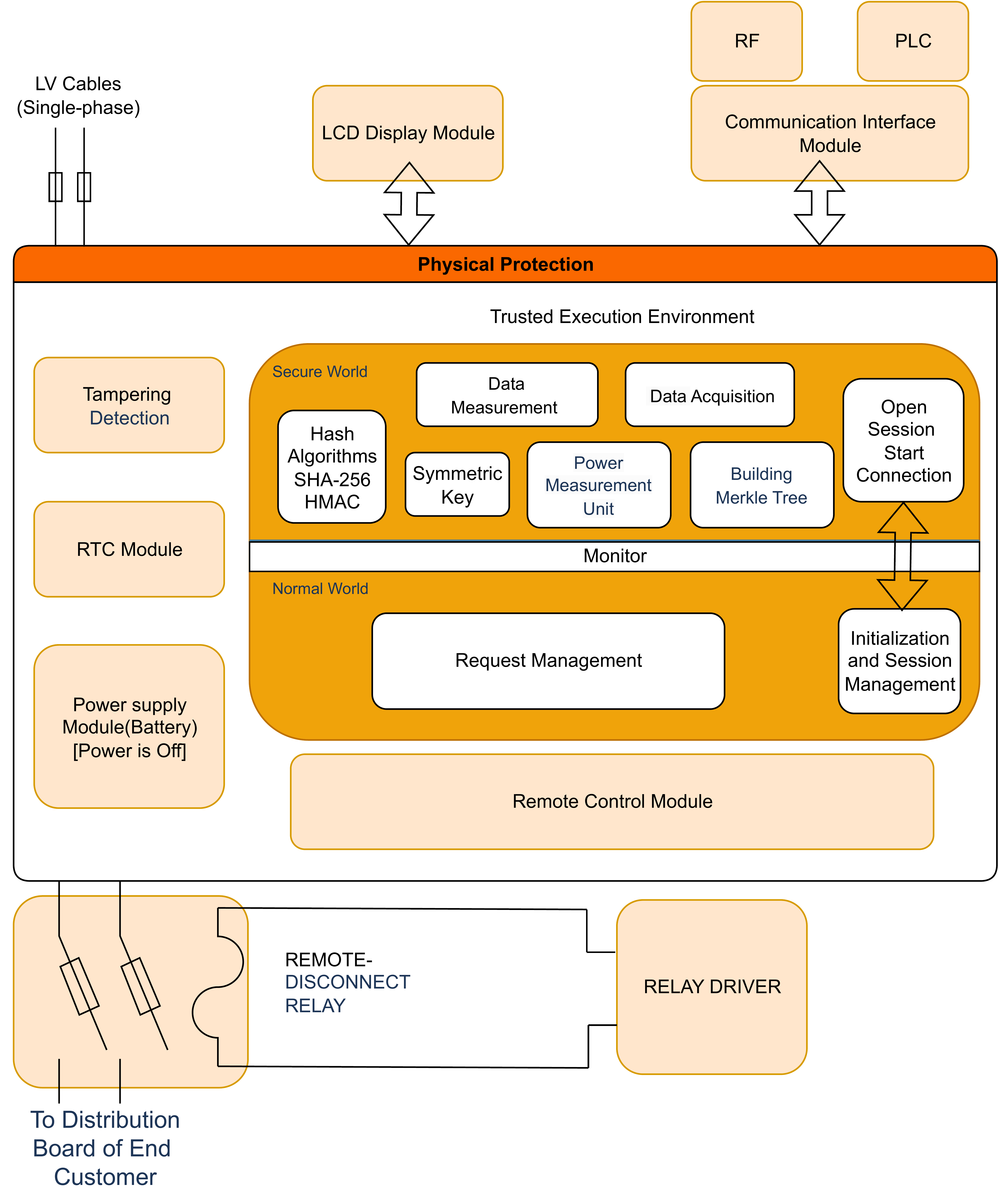}
	\captionof{figure}{Smart meter architecture implemented within a trusted execution environment.}
	\label{smrtmeter}
\end{center}

\begin{itemize}
	\item \textbf{Secure World}
	
	This component is an integrated part of the main processor that performs critical security-related tasks and stores important data. Access to these processes and data requires a series of steps and authentication. that steps unauthorized individuals our subcomponents placed in the trusted execution environment within the meter include:
	\begin{itemize}
		\item\textbf{Hashing:}
		This unit employs SHA-256 and HMAC hash algorithms to secure data transmission, ensuring data integrity and authenticity within the smart meter's Trusted Execution Environment.
		
		\item\textbf{Data Acquisition:}
		This unit is responsible for accurately measuring and monitoring the end user's electricity consumption. It collects real-time energy consumption data and securely processes and transmits it. This ensures accurate billing and correct energy monitoring.
		
		\item\textbf{Data Equalization:}
		This unit ensures the uniformity and consistency of collected data before processing. It harmonizes variations, such as discrepancies in measurement intervals or formats, to enable accurate analysis and secure transmission, essential to maintaining the integrity of the entire data management process.
		
		\item\textbf{Merkle Tree:}
		This unit ensures data integrity by organizing and verifying data through a hierarchical structure. It aggregates hashed data blocks into a tree format, where each leaf node represents a hash of the original smart meter data, and each non-leaf node is the hash of its child nodes. Merkle Tree root derived from this structure produces a single hash value that efficiently summarizes the entire data set, enabling secure and efficient data verification without the need to transmit individual data blocks. This process is critical for secure remote attestation and data authentication.
		
		\item\textbf{Remote Attestation:}
		This unit ensures the integrity and security of the data exchanged between the smart meter and the server. It computes the Merkle Tree root, combines it with a Nonce, and processes it using the HMAC algorithm and a symmetric key to generate a secure hash value. This is sent to the server, allowing the server to verify the authenticity of the data without revealing sensitive information and ensuring secure and trusted communication.                
	\end{itemize}

	\item \textbf{Normal world}
	
	This unit handles and executes processes that do not require security. Access to these processes and data poses no risk, and no special actions are required to access this data at any time.
	\begin{itemize}
		\item\textbf{Monitor(Secure World Handlers):}
		This unit manages the connection between the secure and non-secure worlds without interfering with the generation or processing of sensitive data. Its role is limited to transferring pre-formatted and hashed data, leaving the sensitive data handling and hashing to other dedicated processes.
		
		\item\textbf{Request Management:}
		This unit handles server requests for the smart meter, such as remote attestation or meter update requests. If necessary, it forwards these requests to the monitor to initiate a session within a secure environment.
		
		\item\textbf{Initialization and Session Management:}
		This unit is responsible for monitoring and controlling the secure communication sessions between the secure world and the normal world of the smart meter. It ensures the proper initialization, maintenance, and termination of sessions while maintaining the integrity and confidentiality of data exchanged between these two environments. Managing authentication and access control prevents unauthorized interactions and ensures that sensitive data from the secure world is transmitted and processed securely in the normal world. Thanks to the end-to-end integrity implemented between the secure world and the remote attester, the security of the data is not compromised.
		
	\end{itemize}
\end{itemize}

\begin{figure*}[!htbp]
	\centering
	\includegraphics[width=\textwidth]{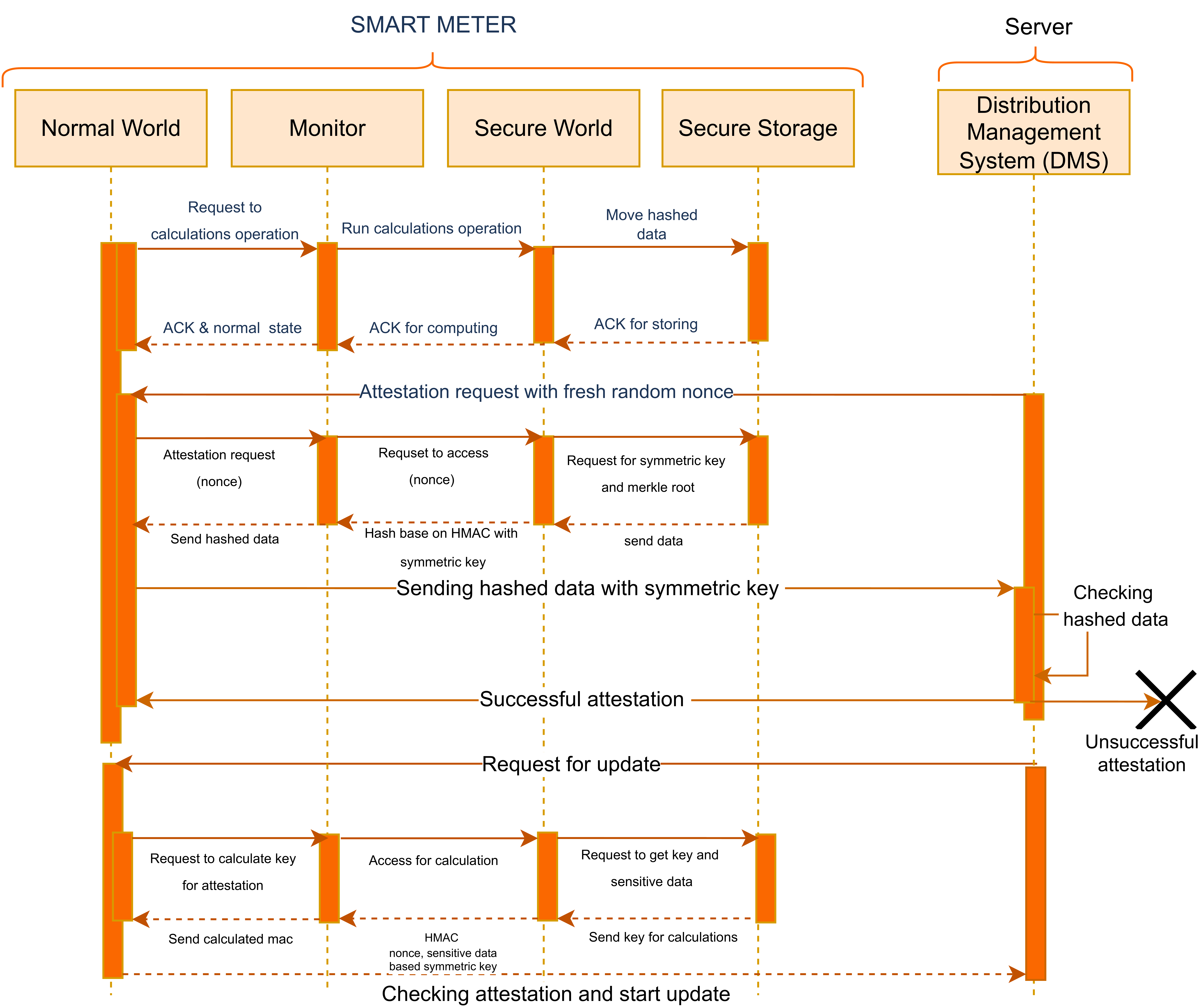}
	\caption{Sequence diagram showcasing the interaction between smart meter components, smart devices, and the control center.}
	\label{seqdiagram}
\end{figure*}

Figure~\ref{seqdiagram} illustrates a smart meter system's secure communication and data management process, highlighting key components such as the normal world, monitor, secure world, secure storage, and the Distribution Management System (DMS). In our approach, remote attestation is initiated to establish a secure connection between the smart meter and the server whenever the metering storage reaches its threshold, ensuring reliable and protected data exchange.

The smart meter communicates with the server to request key calculations for attestation and access and receives acknowledgments (ACKs) for successful updates and data transfers. This systematic process demonstrates a robust approach to managing and transmitting hashed consumption data within a smart meter infrastructure, ensuring data integrity and security through hashing and step-by-step acknowledgments.

\subsection{Implementation}
In this section, we will delve into the details of the implemented scheme. First, we explore the construction of Merkle Tree, followed by an in-depth analysis of the remote attestation process between the server and the smart meter. Afterward, we present the details of the remote attestation.

\subsubsection{Merkle Tree}
The Merkle Tree, a binary tree data structure used for hashing, enables the server to evaluate and validate the basic data critical to the secure operation of the smart meter. A smart meter's initial data — including its unique identifier, security credentials, configurations, manufacturer details, software version, energy consumption profiles, environmental data, communication settings, and activity logs — forms the foundation for the secure and reliable operation of the device. Once connected to the smart grid, this data becomes the leaves of a Merkle Tree, generating a root hash critical for verifying the integrity and authenticity of the meter's data within the grid, ensuring secure and trusted interactions across the grid.

Merkle Tree intermediate computations involve several stages, including:
\begin{itemize}
	\item \textbf{Storing Initial Data:}
	The initialized data $D_{i}$ and critical information within the smart meter are systematically stored in the leaves $L_{i}$ of Merkle Tree, forming a robust and efficient data structure that ensures data integrity and security.
    
	\[\tau_{i} = \text{HMAC}_{k_{i}}(ID_{i} \parallel D_{i} \parallel t)\]
	\[L_{i} = D_{i} \]
    
    where $ID_{i}$ is the factory-assigned device identifier and $t$ is a monotonic counter that binds the tag to a single reporting epoch.

	\item  \textbf{SHA-256:}
	After the data is placed in the leaves of Merkle Tree, it must first be hashed using the SHA-256 hashing algorithm to compute the root of each leaf. After this hashing, the data is placed back into its designated leaf location, ensuring the data structure's security and integrity.
	
	\[H(L_{i}) = \text{SHA\,256}(L_{i})\]
	\[L'_{i} = H(L_{i})\]
	
	\item \textbf{Concatenation:}
	After hashing the individual leaves, the Merkle Tree construction begins by pairing adjacent hashed leaves. Each pair is concatenated by linking the left hashed leaf to the right hashed leaf, forming a unified parent node. This concatenation process is repeated iteratively, building higher levels of the tree, ensuring data integrity, and enabling efficient verification.
	
	\[M_{i, i+1} = L'_{i} \parallel L'_{i+1}\]
	
	\item \textbf{SHA-256 of concatenated leaves:}
	In this step, the two concatenated leaves are subjected to the SHA-256 hashing algorithm. After this process, the middle node of the Merkle Tree is ready, representing a secure and integral part of the tree structure.
	
	\[N_{i, i+1} = \text{SHA\,256}(M_{i, i+1})\]
\end{itemize}

We continue the above steps until the number of intermediate nodes is reduced to two. At this point, we use the same procedure to compute the root of the tree. This root, generated through multiple hashing steps, provides a high-level integrity.
\[M_{Root} = L'_{Left Child} \parallel L'_{Right Child}\]
\[N_{Root} = \text{SHA\,256}(M_{Root})\]

\begin{algorithm}[!htbp]
	\caption{Pseudo-code for the formation of a Merkle Tree and the transfer of its root.} \label{alg:mrk-tree}
	\begin{algorithmic}[1]
		\REQUIRE Initialize LeafNodes, numLeafNodes, merkleRoot
		\ENSURE  The leaf nodes are correctly generated and stored in the LeafNodes array
		\ENSURE The Merkle root is correctly computed and stored in the merkleRoot buffer
		\ENSURE The Merkle-Tree function has been correctly constructed, resulting in a valid Merkle root
		\STATE Iterate through LeafNodes in pairs
		\STATE Calculate hash for each pair of LeafNodes and store it in HashNode array
		\WHILE{number of intermediate nodes > 1}
		\STATE Iterate through intermediate nodes in pairs
		\STATE Calculate hash for each pair of intermediate nodes and store it in HashNode array
		\STATE Reduce the number of intermediate nodes by half
		\ENDWHILE
		\STATE Copy the root hash to merkleRoot and return
		\STATE Return TEE-SUCCESS
	\end{algorithmic}
\end{algorithm}

Algorithm~\ref{alg:mrk-tree} concisely demonstrates Merkle Tree building and root computation in a secure execution environment. The code ensures security by performing root computation and tree construction in a trusted environment. The hashing operations for tree leaves and intermediate nodes are performed in this secure environment using the robust SHA-256 hashing algorithm. Each level of the tree is hashed, significantly reducing the possibility of hostile entities compromising and accessing the original data.


A Merkle Tree is characterized by the cost of three operations: building the tree, verifying a proof, and updating the structure after a leaf changes.

Building the tree is linear in the number of leaves. Hashing proceeds in rounds, and each round produces half as many nodes as the one before it, so the internal nodes account for $n-1$ operations in total, and the construction therefore costs $O(n)$ hash operations for $n$ leaves.

Verification is far cheaper. To decide whether a particular leaf belongs to the tree, one recomputes the hashes along the path from that leaf up to the root, using the sibling values supplied as an authentication path. Since the path is as long as the tree is tall, both the number of hash operations and the size of the proof are $O(\log n)$. It is this
logarithmic behavior that makes Merkle Trees attractive for verifying integrity over large datasets.
Updates share the same logarithmic $O(\log n)$ cost instead of a full rebuild, and any proof from an out-dated tree will fail against the new root.

\subsubsection {Remote Attestation}
\begin{figure*}[!htbp]
	\centering
	\includegraphics[width=\textwidth]{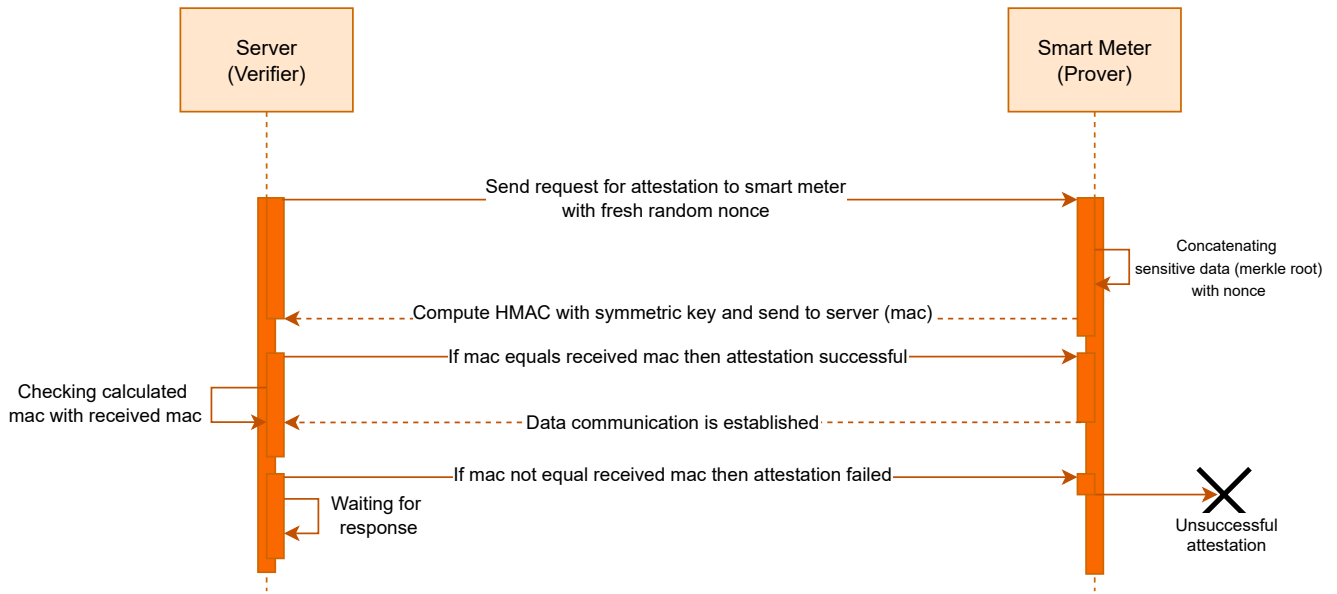}
	\caption{Sequence diagram of the remote attestation process in the implemented model.}
	\label{ra-fig}
\end{figure*}

Remote attestation is a process that involves two entities: the Verifier and the Prover.
In this context, remote attestation attempts to verify the integrity of the smart meter under investigation.
In Figure~\ref{ra-fig}, we present the sequence diagram of the remote attestation between the smart meter and the server in our implemented method, which includes the following steps:

\begin{itemize}
	\item \textbf{Remote Attestation:}
	Initially, the server establishes a connection with the smart meter by sending a fresh random Nonce $N$ and a timestamp $T_s$, all hashed by the pre-shared symmetric key $K$ to initiate the communication process.

	\begin{gather*}
		\text{Server} \rightarrow \text{Smart Meter}: \text{New Attestation}, \\
		N, T_s,  \text{HMAC} (K, N || T_s)
	\end{gather*}

	\item \textbf{Data calculation:}
	The smart meter checks if $T_s$ is current and $HMAC$ is valid, then records $T_s$. If another request with $T_s$ is received, it is rejected, and the smart meter concatenates the root of the Merkle Tree $N_{root}$ with the received fresh random Nonce value $N$.
	
	\[M = N_{root} \parallel N\]
	
	\item \textbf{HMAC:}
	It then applies the $HMAC$ hashing algorithm with key $K$ on $M$ and transmits the hashed value $H$ along with the Nonce $N$ back to the server.
    
	\[ H = $HMAC$ (K, M) \]
	\[Smart Meter \rightarrow Server:(H,N)\]
	
	\item \textbf{Communication}
	The server recalculates the hash using its store root of Merkle-tree $N{root}$, Nonce $N$, and its symmetric key and compares it to the received data. The remote attestation succeeds if the data is correctly hashed and has not been tampered with during transmission. The remote attestation will fail if the data is not valid. 
	
	\[M' = N{root} \parallel N\]
	\[H' = $HMAC$ (K, M') \]
	
	\item \textbf{Successful Attestation:}
	Once remote attestation is successfully established between the smart meter and the server, a secure session is initiated to facilitate the transfer of sensitive data, including measurement and privacy-related information. All data exchanges between the meter and the server occur within this secure session. Additionally, if the smart meter requires an update, the server utilizes the established secure session to transmit the updated data securely. After the update is applied, remote attestation is performed again to verify that the update has been successfully executed.
	
	\item \textbf{Unsuccessful Attestation:}
	If the hashed value in the smart meter does not match the value stored on the server, the remote attestation between the smart meter and the server will fail. Upon a failed remote attestation, the meter will be added to the server’s blacklist. If subsequent remote attestations also fail after several attempts at different time intervals, the electricity distribution company will inspect the problematic meter to identify and resolve any underlying issues.
	
\end{itemize}

\begin{algorithm}[!htbp]
	\caption{Pseudo-code for remote attestation between smart meter and server.} \label{alg:RA-sm-sr}
	\begin{algorithmic}[1]
		\REQUIRE Make a Nonce, Initialize smart meter, Initialize server 
		\ENSURE The server establishes a connection with the smart meter by transmitting a fresh random Nonce
		\ENSURE The Smart Meter sends an HMAC calculated over the concatenated Nonce and sensitive data to the Server
		\STATE Server Generates a Challenge Based on Nonce 
		\STATE Server Sends Challenge to Smart Meter
		\STATE Smart Meter Generates an HMAC based on Nonce and sensitive data and symmetric key
		\STATE Smart Meter Sends Response to Server
		\STATE Server calculates and Verifies the Response based on the symmetric key and stored root
		\IF{Response is valid and equal with received HMAC}
		\STATE Attestation Successful
		\ELSE
		\STATE Attestation Failed
		\ENDIF 
	\end{algorithmic}
\end{algorithm}

Algorithm~\ref{alg:RA-sm-sr} outlines a remote attestation process between a smart meter and a server to ensure data integrity and authenticity.

Remote attestation of smart meters is conducted periodically, aligning with scheduled data transmissions to the server rather than being performed continuously. This approach establishes a secure connection between the server and the meter and utilizes the Trusted Execution Environment (TEE) to securely store read data until transmission occurs. By maintaining data integrity and protection within the TEE until the successful execution of remote attestation, the system ensures that the data remains secure during transit. Upon successful transmission to the server, the stored data is replaced with newly read data. According to the electricity company's policies, the implementation of remote attestation is optimized to minimize overhead on the smart grid. Since data transfer from smart meters to the server is a routine operation, and the runtime and frequency of remote attestation executions are carefully managed, the overall impact on system performance remains negligible. This ensures that enhanced security measures do not compromise the efficiency and reliability of the smart grid infrastructure.

\subsection{Key Management}

Effective key management is central to the security guarantees of the proposed architecture. Given the resource-constrained nature of the smart meter platform---an ARM Cortex-M33 based microcontroller with limited processing power, memory, and energy---the design deliberately relies on symmetric key cryptography rather than asymmetric (public-key) schemes for protecting communication between the meter and the Distribution Management System (DMS).

\subsubsection{Rationale for Symmetric Key Cryptography}

Asymmetric cryptographic algorithms, such as RSA or elliptic-curve schemes, typically require significantly larger key sizes, more complex arithmetic operations, and greater memory footprint for both code and intermediate computation. On a resource-constrained embedded device such as the smart meter used in this work, these requirements translate directly into higher energy consumption, increased latency per cryptographic operation, and additional RAM overhead that competes with the already limited resources available inside the Secure World.

By contrast, symmetric key cryptography---and in particular HMAC-based constructions---offers substantially lower computational cost, smaller code size, and faster execution, all of which are critical for a device that must perform frequent metering, billing, and attestation operations under tight power and memory constraints. This efficiency advantage, combined with the fact that the TEE already provides a hardware-isolated environment for secure key storage (eliminating much of the motivation for public-key based key exchange), makes symmetric cryptography the more practical and economically justified choice for this class of critical devices.

\subsubsection{Secure Key Storage within the TEE}

Rather than relying on an external Secure Element (SE) or Hardware Security Module (HSM), the symmetric key used by each meter is generated and stored entirely within the Secure Storage component of the on-chip Trusted Execution Environment (TrustZone-M Secure World). Because this key resides at the silicon level---physically isolated from the Normal World and inaccessible to the untrusted application code---it cannot be extracted, traced, or tampered with through software-based attacks originating outside the Secure World. This hardware-enforced isolation removes the need for additional discrete security hardware, directly reducing the Bill of Materials (BoM) while preserving a level of key protection comparable to that of a dedicated SE/HSM.

The key never leaves the Secure World during normal operation: all operations that require the key---hashing, HMAC computation, and verification---are performed internally by the Secure World, and only the resulting output (e.g., a hashed value) is passed to the Normal World or transmitted externally.

\subsubsection{Key Usage during Remote Attestation}

The symmetric key is central to the remote attestation protocol shown in Figure~\ref{seqdiagram}. 

When the Distribution Management System sends an attestation request with a fresh random nonce, the meter's Secure World hashes the relevant sensitive data using a Merkle tree and generates only the root for transmission, which also keeps communication overhead low—a helpful property for a bandwidth- and energy-constrained device.

This root is then authenticated using HMAC with the device's symmetric key, producing a keyed hash that binds the attestation response to both the fresh nonce and the specific device key. The resulting HMAC value, together with the hashed data, is transmitted to the DMS, which independently verifies the value using its own copy of the corresponding key material. A successful match confirms that the data originated from a genuine, uncompromised device; any mismatch causes the attestation to fail.

Put simply, HMAC provides integrity and authenticity for the attestation response. Confidentiality is not required for the attestation metadata in the present design; however, if a deployment requires confidential attestation reports, an authenticated-encryption mechanism can be incorporated. There is no need for the heavier asymmetric key exchange, and the hardware isolation offered by the TEE is leveraged.

\subsubsection{Rationale for Symmetric Attestation Keys}

The selection of symmetric attestation keys follows from the operating regime of the target platform, in which a fresh tag is produced on every reporting interval and per-attestation cost is therefore incurred
continuously. A keyed hash requires a small constant number of compression-function calls and a single block of working state, which keeps the secure-world footprint at 2760 bytes of static and 22564 bytes
of dynamic RAM even for a 512-leaf tree
(Section~\ref{subsec:ram}), leaving headroom within the memory budget of the metering hardware. Certificate-based and blockchain-based attestation offer complementary properties, notably public verifiability and
independence from a shared secret held by the verifier.
These solutions are well suited to settings where endpoints can accommodate asymmetric arithmetic and, in the case of a blockchain, the associated communication
latency and transaction overhead. The present design targets the opposite end of that spectrum, favouring a minimal on-device trusted computing base and a smaller overhead per attestation. The corresponding
trade-off is that complexity moves from computation to key management, which is addressed in
Section~\ref{subsec:key-lifecycle}.
In addition, each key is unique to a single device and never leaves the secure world, hence, the effect of a compromise remains confined to that device.

\section{Evaluation}
\label{sec:evaluation}

In this section, we begin by reviewing and evaluating the trusted execution environment provided by OP-TEE, followed by a comprehensive comparison of the proposed method with related works, analyzing various aspects of their performance and applicability. Next, we assess the performance of the proposed method by examining key metrics such as runtime and memory consumption, with particular attention to the resource requirements of Merkle Tree leaf nodes. Subsequently, we conduct a Failure Resistance Analysis to evaluate the system’s robustness against potential failure scenarios. Finally, we analyze Adversary and Attack Scenarios to assess the method’s resilience to security threats.

\subsection{Security Comparison}
We conduct a comprehensive comparison of the proposed method with similar approaches, examining different parameters and perspectives.
In particular, we focus on Merkel tree leaves and investigate execution times for different types of smart meters.

The Unified CS-based Framework~\cite{lee2019unified} integrates compression and authentication within Advanced Metering Infrastructure (AMI). Utilizing compressive sensing (CS) with residual error-based hypothesis testing, this framework achieves energy efficiency, reduced computational complexity, and low authentication error rates. However, it faces notable challenges, such as potential fidelity loss due to lossy compression, limited adaptability to diverse attack scenarios, and deployment complexity in real-world environments. In contrast, our proposed remote attestation method leverages a Merkle Tree structure within a Trusted Execution Environment (TEE) using OP-TEE, ensuring robust data integrity and privacy without compromising data fidelity. Unlike the Unified CS-based Framework, our approach avoids lossy compression, thereby maintaining the accuracy of energy consumption data. Additionally, the implementation within a secure TEE simplifies deployment complexities and enhances adaptability to various attack vectors, making our solution more resilient and easier to integrate into existing smart grid systems.

Logical Key Hierarchy (LKH)~\cite{abdalzaher2023survey} is recognized for its scalable key management utilizing a tree-based structure in dynamic smart grid environments. However, the framework encounters significant drawbacks, including resource-intensive rekeying processes during node compromises and increased complexity in multicast group management. Our method addresses these limitations by eliminating the need for complex key hierarchies through Merkle Trees for data integrity and authentication. The Merkle Tree approach inherently provides a scalable verification mechanism without the overhead associated with key management in LKH. Furthermore, executing all critical operations within a TEE ensures that key management and data authentication are securely handled, reducing the risk of resource exhaustion and simplifying the overall security architecture.

\begin{table*}[!t]
	\centering
	\caption{Comparison of smart meter security methods.}
	\label{tab:comparison}
    \setlength{\tabcolsep}{3pt}

\begin{tabularx}{\textwidth}{|p{2.4cm}|p{0.8cm}|X|p{1.3cm}|p{1.3cm}|p{1.3cm}|p{1.3cm}|}		
\hline
		\textbf{Work} & \textbf{Year} & \textbf{Disadvantages} & \textbf{Resource Efficiency} & \textbf{Ease of Deployment} & \textbf{Scalability} & \textbf{Security Robustness} \\ 
		\hline
		Unified CS-based Framework~\cite{lee2019unified} & 2019 & Potential fidelity loss, limited adaptability to diverse attacks, deployment complexity. & Medium & Medium & Medium & Medium \\ 
		\hline
		Logical Key Hierarchy (LKH)~\cite{abdalzaher2023survey} & 2023 & Resource-intensive rekeying during node compromise, multicast group complexity. & Low & Low & High & Medium \\ 
		\hline
		Physically Unclonable Functions (PUFs)~\cite{al2023physical} & 2023 & High cost, real-time computational demands, integration challenges. & Low & Low & Medium & High \\ 
		\hline
		Cloud-Edge-End Architecture~\cite{jiang2020lightweight} & 2020 & Computational overhead, deployment complexity for low-cost meters. & Low & Low & High & Medium \\ 
		\hline
		EPDA Scheme~\cite{alsharif2019epda} & 2019 & Infrastructure integration challenges, scalability issues in large-scale deployments. & Medium & Low & Medium & High \\ 
		\hline
		P2P Energy Theft Detection~\cite{salinas2013privacy} & 2013 & Numerical instability in large networks, limitations in decomposition accuracy. & Medium & High & Medium & Medium \\ 
		\hline
		State Estimation with PPBE~\cite{salinas2015privacy} & 2015 & Computational complexity, reliance on accurate system parameters. & Low & Medium & High & Medium \\ 
		\hline
		\textbf{Proposed Method} & 2026 & Requires TEE Support & High & High & High & High \\ 
		\hline
	\end{tabularx}
\end{table*}

Physically Unclonable Functions (PUFs)~\cite{al2023physical} offer hardware-oriented authentication by generating unique cryptographic keys from inherent hardware randomness, providing resistance against impersonation attacks and reducing data transmission overhead. Despite these strengths, PUF-based systems are often hindered by high costs, real-time computational demands, and integration challenges within existing AMI infrastructures. Our proposed authentication method mitigates these issues by utilizing a Merkle Tree within a secure TEE, which is both cost-effective and computationally efficient. Our approach can be implemented on standard smart meters with minimal modifications. Additionally, the secure execution environment of OP-TEE ensures that authentication processes are both swift and reliable, making our solution more practical for widespread deployment in resource-constrained smart grid systems.

The Cloud-Edge-End Architecture~\cite{jiang2020lightweight} enhances AMI security by integrating localized edge computing for data processing and employing Fully Homomorphic Encryption (FHE) for secure data aggregation. This architecture achieves low latency and high bandwidth utilization while enabling secure encrypted data operations. However, it introduces significant computational overhead and deployment complexity, particularly for low-cost smart meters with limited resources. In comparison, our remote attestation method offers a streamlined approach focusing on efficient data integrity verification using Merkle Trees within a TEE. This eliminates the need for computationally intensive encryption processes like FHE, thereby reducing the overall computational burden. Furthermore, our solution simplifies deployment by leveraging existing trusted execution environments, making it more suitable for low-cost smart meters without compromising security or performance.

The Efficient and Privacy-Preserving Data Aggregation (EPDA)~\cite{alsharif2019epda} scheme addresses multi-recipient AMI networks by employing proxy re-encryption mechanisms to facilitate secure data aggregation and access control. While EPDA effectively resists collusion attacks and ensures fine-grained access control with low computational overhead, it requires substantial infrastructure modifications and complex key management systems for integration into existing deployments. Our proposed method surpasses EPDA by offering a more straightforward integration process. By centralizing the integrity and authenticity verification within a secure environment, our approach reduces the need for extensive infrastructure changes and simplifies key management. Additionally, Merkle Trees’ efficient construction and verification ensure that our method maintains low computational and memory overhead, making it more adaptable to large-scale and resource-constrained AMI deployments.

Peer-to-Peer (P2P) Energy Theft Detection~\cite{salinas2013privacy} utilizes distributed algorithms like LU and QR decomposition to identify fraudulent users without compromising individual privacy. While this method offers privacy protection and scalability, it is susceptible to numerical instability in large networks and limitations in decomposition accuracy. Our approach enhances energy theft detection by providing a more stable and accurate verification mechanism. The secure execution environment ensures that all data integrity checks are performed reliably, mitigating the risks of numerical instability. Moreover, using Merkle Trees allows for efficient and scalable data verification, which is crucial for large-scale smart grid systems. This results in a more dependable and secure framework for detecting energy theft while preserving user privacy.

The Privacy-Preserving Bias Estimation (PPBE) framework~\cite{salinas2015privacy} employs Kalman filters for state estimation-based energy theft detection. However, it suffers from high computational complexity and dependency on accurate system parameters, which can be challenging to maintain in dynamic smart grid environments. In contrast, our proposed method leverages Merkle Trees within a TEE to achieve efficient and reliable data integrity verification with significantly lower computational demands. By simplifying the state estimation process and reducing dependency on precise system parameters, our approach offers a more scalable and adaptable solution for energy theft detection. Additionally, the secure environment provided by OP-TEE ensures that all verification processes are both secure and efficient, enhancing the overall reliability of the smart grid system.

As demonstrated in Table~\ref{tab:comparison}, our proposed remote attestation method significantly outperforms existing smart meter security frameworks across all evaluated criteria, including Resource Efficiency, Ease of Deployment, Scalability, and Security Robustness. Unlike the Unified CS-based Framework and Logical Key Hierarchy (LKH), which achieve only medium performance and encounter challenges such as deployment complexity and limited adaptability, our method consistently scores high, ensuring efficient data verification and seamless integration into existing smart grid systems. Furthermore, while Physically Unclonable Functions (PUFs) and the EPDA Scheme offer robust security features, they are hindered by high costs and integration difficulties, respectively. In contrast, our approach provides a cost-effective and easily deployable solution without compromising security or scalability. The Cloud-Edge-End Architecture and P2P Energy Theft Detection methods, although scalable, suffer from computational overhead and numerical instability, issues that our method effectively mitigates. Overall, Table~\ref{tab:comparison} underscores the advantages of our proposed method, positioning it as a superior and practical solution for enhancing the security and reliability of smart grid systems. 

A notable disadvantage of our approach is its reliance on Trusted Execution Environment (TEE) support, which may limit deployment in systems without TEE capabilities or increase hardware and implementation costs.

In other words, our proposal is intended for next-generation smart meters, new deployments, and modernization programs based on TEE-capable hardware, rather than as a transparent software retrofit for all deployed legacy meters.

\subsection{OP-TEE Environment}
The Open Portable Trusted Execution Environment (OP-TEE) is an open-source TEE widely used to develop TEE-based solutions. While OP-TEE is primarily based on Arm Trust Zone technology, it implements Global Platform’s TEE Internal Core API v1.3.1 specification, making it compatible with other TEE solutions. We use OP-TEE in a virtual Arm environment for our implementation portability, which allows OP-TEE to function across different architectures and hardware configurations, supporting multiple operating systems and trusted execution environments. As an open-source platform, OP-TEE offers transparency, customization, and ongoing development by a broad community of developers. Its compatibility with ARM processors equipped with TrustZone technology allows for broad deployment across various devices, while its adherence to the GlobalPlatform APIs ensures interoperability with secure applications. In addition, OP-TEE emphasizes security through features such as secure boot, isolated execution, and efficient memory management, making it an ideal solution for embedded systems that require a compact, reliable, and versatile execution environment.

\subsection{Run Time Evaluation}
The OP-TEE secure environment was set up on a virtual machine using Oracle VM VirtualBox Manager to conduct the performance analysis. This environment was designed based on ARM v8.0 architecture and optimized for STM32L5 hardware.
Ubuntu 18.04 was used to configure the virtual environment, with 8 GB of dedicated memory and eight virtual processors allocated.
The host system supporting this virtual environment featured 16 GB of RAM and an Intel® Core™ i7-7700HQ CPU @ 2.80 GHz, running Windows 11 Pro.

Throughout this analysis we assume that the verification server is provisioned with sufficient computing and storage resources for the attestation workload, as is typical for utility-side infrastructure. The assumption is an analytical one: it isolates the cost of the security mechanisms themselves---remote attestation and Merkle tree construction---from server-side hardware constraints, so that the measured execution times reflect computational complexity and data volume rather than provisioning choices. Under concurrent load, the aggregate verification latency at the server grows linearly with the number of reporting meters.

Figure~\ref{fig:runtimecmp} clearly shows the direct relationship between the execution time of the smart meter remote attestation process, which includes Merkle Tree as a key component, and the size of Merkle Tree, defined by the number of leaves. The horizontal axis of the figure represents the number of Merkle Tree leaves - 128, 256, 512, and 1024 - while the vertical axis shows the corresponding execution time in seconds. The execution time increases from 3.898 to 27.077 seconds as the number of leaves increases from 128 to 1024.

\begin{center}
	\includegraphics[width=0.95\columnwidth,height=0.7\textheight,keepaspectratio]{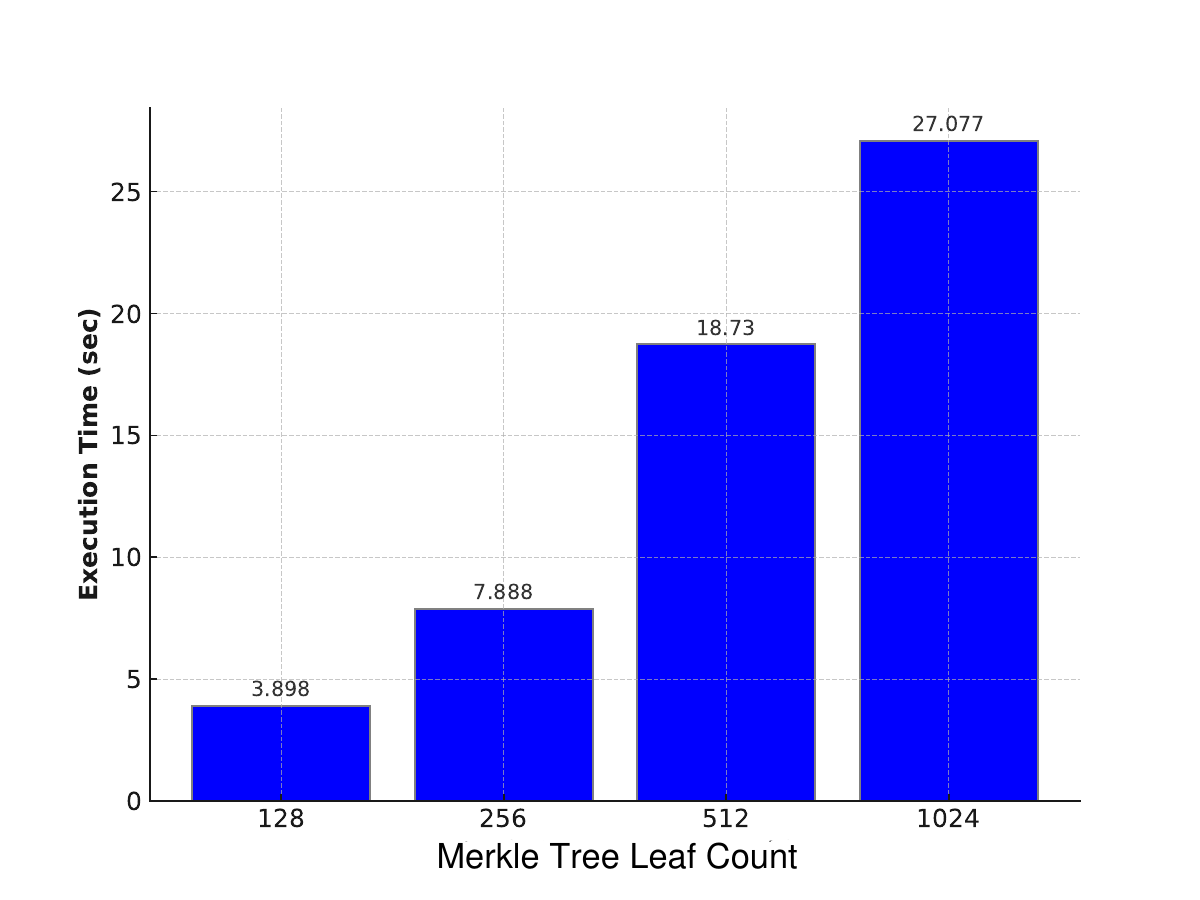}
	\captionof{figure}{Comparison of run time for various numbers of Merkle Tree leaves in a secure smart meter.}
	\label{fig:runtimecmp}
\end{center}

This increase in execution time is due to the increased computational complexity involved in processing and verifying larger Merkle Trees. This consideration is critical in the design and implementation of remote attestation-based systems, as selecting the appropriate Merkle Tree size can significantly impact system efficiency and response time. Thus, in practical applications where execution time is crucial, this parameter must be carefully considered to strike an optimal balance between security and system performance.


Communication security between smart meters and servers is critical due to the exchange of sensitive information. Processes such as remote attestation and various cryptographic techniques are used to secure these communications. This study investigates the performance and execution time of key components within the remote attestation process simulated in the OP-TEE environment. The components analyzed play a critical role in ensuring the security and integrity of transmitted data, including Open Session, Random Number Generation, HMAC, SHA-256, and Remote Attestation. Detailed execution times for each process are meticulously analyzed in Figure~\ref{fig:partstime}, which clearly illustrates the impact of each component, both individually and in conjunction with others, on overall system performance. This analysis underscores the importance of optimizing the remote attestation process to balance security and system efficiency.


\begin{center}
	\includegraphics[width=0.95\columnwidth,height=0.7\textheight,keepaspectratio]{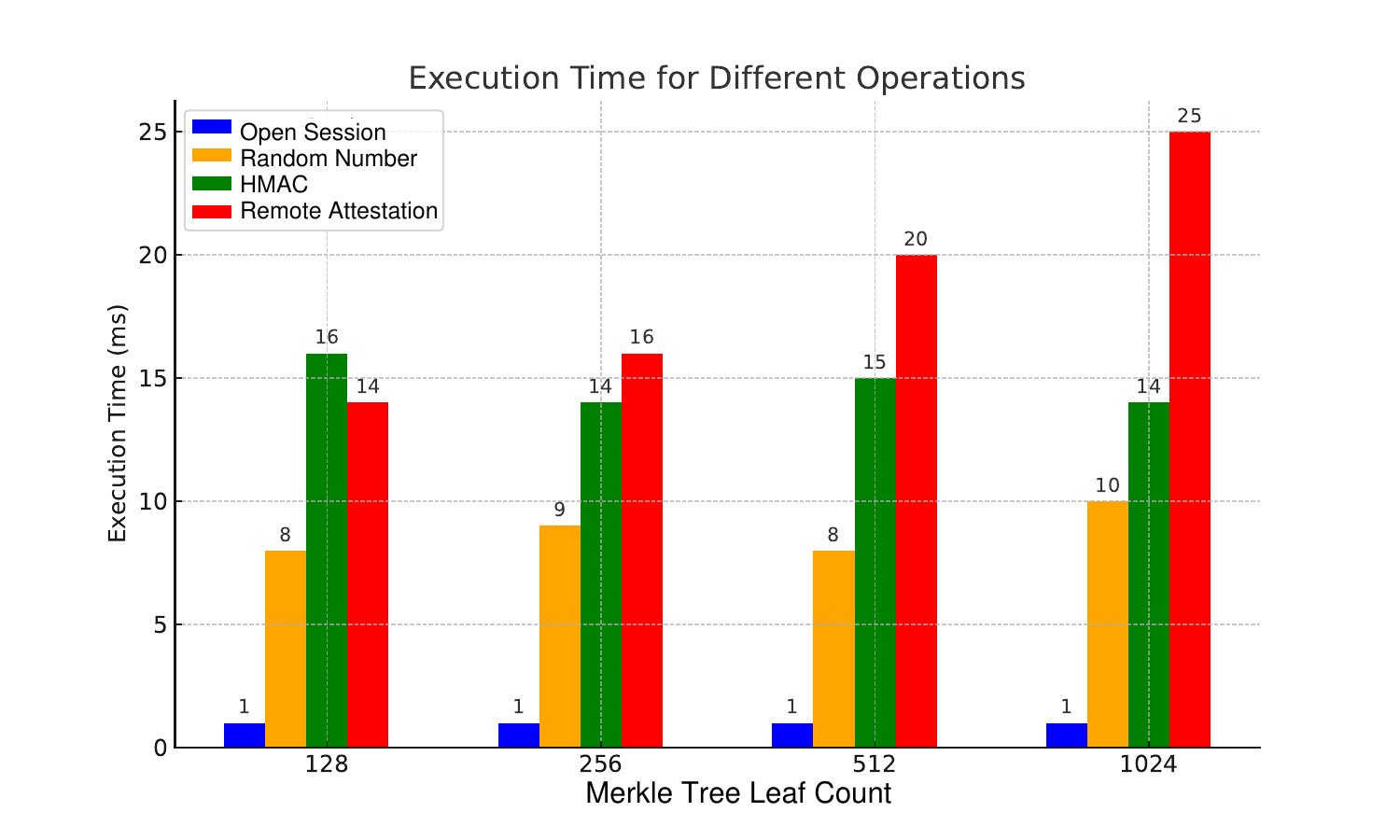}
	\captionof{figure}{The execution time of remote attestation components in the simulated environment.}
	\label{fig:partstime}
\end{center}

In our runtime evaluation, we analyzed the execution time of various components as the number of leaves in the Merkle Tree increased from 128 to 1,024. We optimized the algorithm to address the escalating computational complexity associated with larger Merkle Trees, ensuring it remains efficient as the tree size grows. Conversely, components such as session establishment, random number generation, and HMAC hashing maintained constant runtimes. This consistency arises because these components are independent of the Merkle Tree size and are exclusively utilized in the remote attestation process that leverages the Merkle root. Our results highlight the critical importance of optimizing remote attestation processes to balance security measures and overall system efficiency effectively. By ensuring that the core components efficiently scale while maintaining essential security functionalities, our approach demonstrates both practicality and scalability for deployment in resource-constrained environments like smart meters.

Figure~\ref{fig:MerkleTime} illustrates the execution times for Merkle Tree construction and computation within OP-TEE, divided into two key stages. The first phase, which occurs in the secure world, involves the initial construction of the tree, including hash computations and the integration of leaves to generate the root. As expected, the execution time in this phase scales with the number of leaves, increasing from approximately 328 milliseconds for 128 leaves to 5661 milliseconds for 1024 leaves. This growth reflects the increasing computational complexity required for larger Merkle Trees. The second stage involves interaction with secure storage, in which information about the Merkle Tree leaves is exchanged. Similarly, execution times at this stage increase with the number of leaves, from 1789 milliseconds for 128 leaves to 21242 milliseconds for 1024 leaves. These results provide a detailed insight into the performance overhead imposed by different Merkle Tree sizes in the OP-TEE environment and highlight the relationship between tree size and computational and communication requirements.


\begin{center}
	\includegraphics[width=0.95\columnwidth,height=0.7\textheight,keepaspectratio]{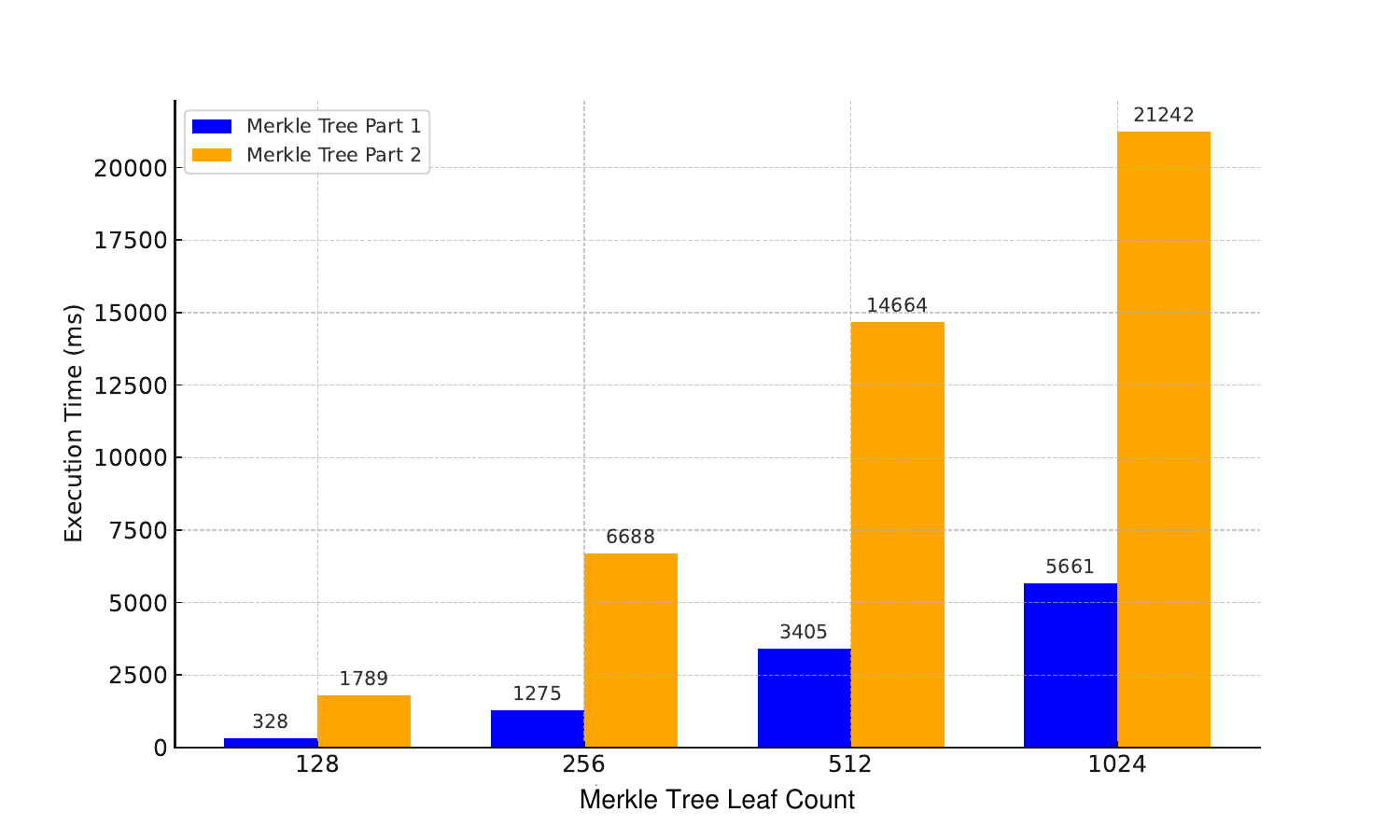}
	\captionof{figure}{Execution time of Merkle Tree formation in the simulated environment.}
	\label{fig:MerkleTime}
\end{center}


\subsection{Memory Usage Evaluation}
\label{subsec:ram}
In this section, we analyze the RAM consumption of the smart meter. As shown in Figure~\ref{fig:totalRam}, the analysis focuses on RAM usage while implementing a Merkle Tree with 512 leaves. This examination provides a detailed insight into the memory requirements associated with constructing and managing the Merkle Tree structure.

The RAM consumption for the proposed method is divided into two types: static RAM and dynamic RAM. The static RAM consumption is reported to be 2760 bytes, while the dynamic RAM consumption is 22564 bytes. The results indicate that the dynamic part of the system requires more memory due to the more complex processes associated with the Merkle Tree within the Trusted Execution Environment. However, the overall RAM consumption remains within limits that do not negatively impact the system’s overall performance, maintaining a balanced trade-off between security and efficiency. Note that we consider a large Merkle Tree for scalability, while the traditional smart meters do not require such a sophisticated tree.


\begin{center}
	\includegraphics[width=0.95\columnwidth,height=0.7\textheight,keepaspectratio]{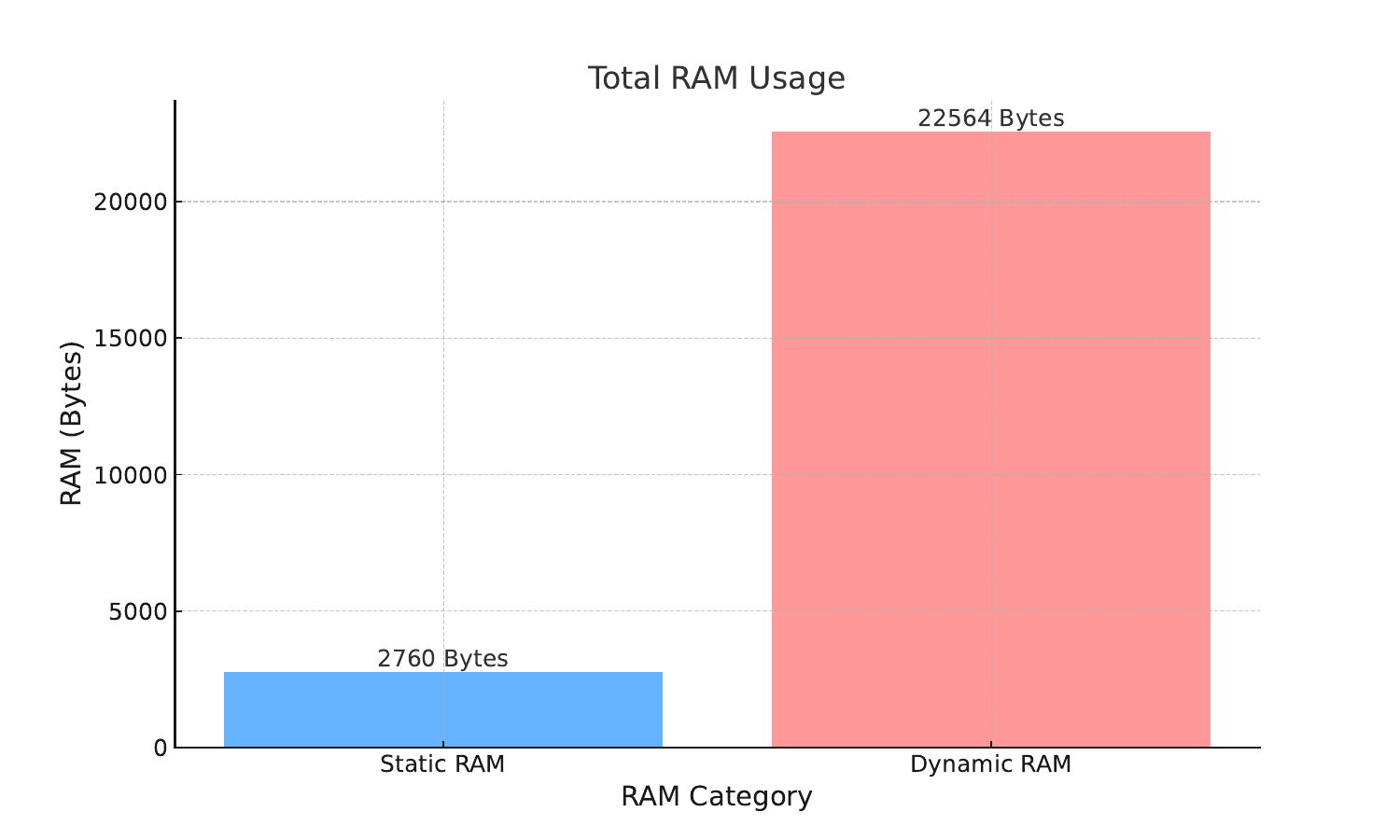}
	\captionof{figure}{Dynamic and static RAM consumption during remote attestation execution.}
	\label{fig:totalRam}
\end{center}

\begin{figure*}[!htbp]
	\centering
	\includegraphics[width=\textwidth]{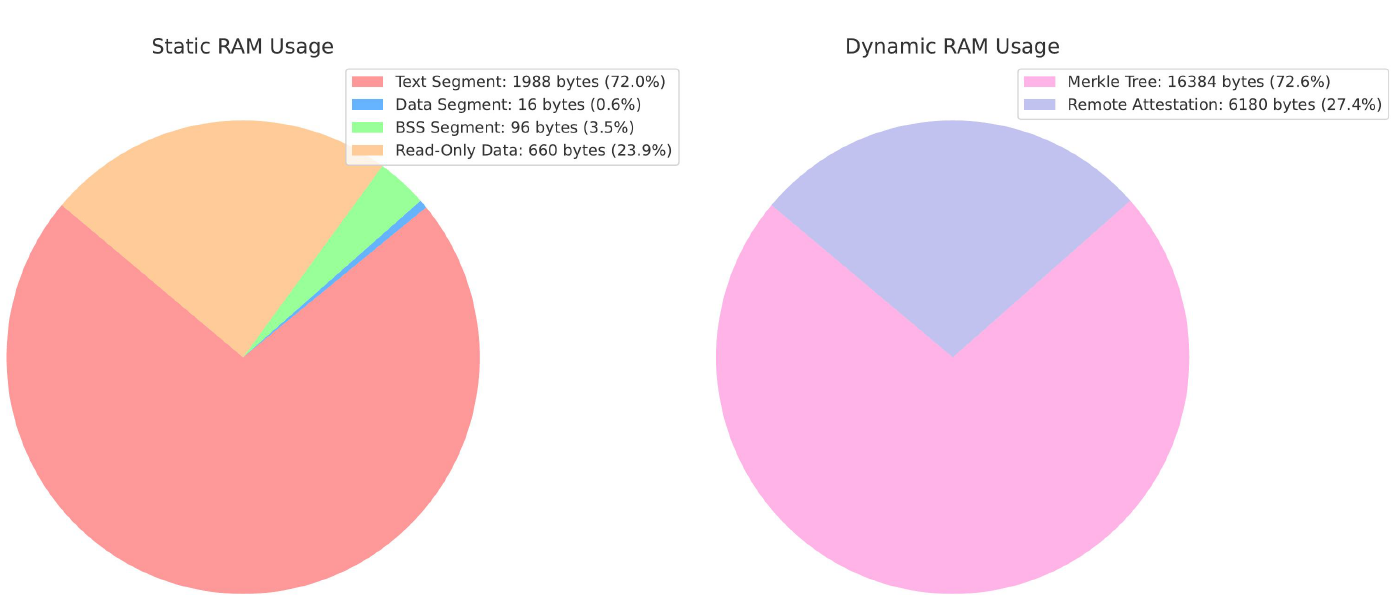}
	\caption{Dynamic and static RAM consumption during remote attestation execution.}
	\label{fig:detailRam}
\end{figure*}

Figure~\ref{fig:detailRam} examines the details of static and dynamic RAM consumption within the TEE in different parts of the system, highlighting the contribution of each component to the overall RAM consumption.

The total consumption in the static RAM section is divided into four parts. The text segment, which consumes 1988 bytes (72.0\%), has the largest share, representing the executable code used during program execution. The Read-Only Data section is the second largest consumer with 660 bytes (23.9\%) and contains data that remains unchanged during program execution. The BSS segment, with 96 bytes (3.5\%), contains uninitialized data, while the Data segment, with 16 bytes (0.6\%), contains initialized, writable data.

In the dynamic RAM area, two main components are the primary consumers. The Merkle Tree is the largest consumer of dynamic RAM, consuming 16384 bytes (72.6\%). This wage is directly related to the tree size of 1024 nodes (512 leaves). The Remote attestation process, which consumes 6180 bytes (24.7\%), is the other primary dynamic RAM user dedicated to ensuring the security of remote communications.

This analysis demonstrates that the Trusted Execution Environment (TEE) effectively manages both static and dynamic RAM consumption, providing high security without imposing excessive load on system resources. Each system component contributes to overall security, with RAM usage proportional to its specific role. These findings underscore the importance of resource allocation and optimization in designing secure systems and can serve as a foundation for further research and enhancements in this field.

\subsection{Failure Resistance Analysis}
In this section, we evaluate the system’s resilience against various failure modes affecting different smart grid components.

\subsubsection{Power Outage Scenario}
In the event of a complete power outage across the smart grid, integrating a battery and selecting non-volatile storage within the smart meter ensure uninterrupted data storage. The critical meter information is securely saved and preserved, preventing data loss during the outage. Upon power restoration, the system seamlessly reestablishes a remote attestation between the server and the smart meter. This process guarantees the continued integrity and reliability of the smart grid by ensuring that all previously stored data remains intact and trustworthy.

\subsubsection{Communication Disruption Scenario}
Another critical failure mode involves the disconnection of the communication network between the smart meter and the server. Thanks to the implementation of a TEE within the smart meter, measured data remains secure. The TEE operates in a secure, isolated space, protecting the data from unauthorized access or tampering, safeguarding data integrity, and maintaining privacy. Eventually, when the network connection is restored, the meter updates the server with the measurements.

These failure resistance mechanisms collectively establish a comprehensive and resilient framework for managing smart meter data securely. By addressing both power and communication failures, the system ensures continuous operation and data integrity, even under adverse conditions. This robustness not only enhances the reliability of individual smart meters but also fortifies the overall security and efficiency of the entire smart grid infrastructure.

\subsection{Adversary and Attack Scenarios}
As critical components of smart grids, smart meters measure and transmit sensitive household energy consumption data, making them a potential target for malicious adversaries. This section outlines the adversarial model, capturing the capabilities and intentions of attackers who aim to compromise the integrity, authenticity, and privacy of smart meter data. By leveraging a Merkle Tree-based authentication method and the secure environment of a TEE, the proposed system addresses threats such as data tampering, forgery, and replay attacks. The model also considers advanced threats targeting the TEE and ensures robust countermeasures to maintain secure communication and reliable data verification, enhancing the overall security of the smart grid infrastructure.

\subsubsection{Man-in-the-Middle}
The proposed remote attestation method for smart meters mitigates the risk of Man-in-the-Middle (MITM) attacks by utilizing a pre-shared symmetric key between each meter and the server. The key is fed to the HMAC
operations so that only the authentic meter and server can generate a correct message. It is provisioned at manufacturing time and held in the secure storage of the TEE, and it is never exported to the normal
world. In an MITM attack, an adversary intercepts communication between a smart meter and the central server, potentially altering or forging messages to compromise data integrity. The method addresses this by
organizing transmitted data into a Merkle Tree, where the root hash acts as a compact and immutable representation of the entire dataset. Any modification to the transmitted data during an attack results in a mismatch
in the Merkle root, making tampering immediately detectable during verification. Without the per-device key, an adversary cannot construct a leaf that validates against the published root, and the freshness input
included in each tag prevents a previously observed message from being replayed in a later epoch. Should a device be compromised, the device-specific key is blacklisted by the DMS, so that any future
messages are rejected. Additionally, all operations involved in computing the Merkle root and verifying hashes are performed within the TEE, ensuring the integrity of
the process. These hash-based mechanisms, combined with the TEE's secure execution, prevent adversaries from injecting or modifying data undetected, thereby safeguarding the authenticity and integrity of smart meter
communications against MITM attacks.

\subsubsection{Replay Attack}
The proposed remote attestation method for smart meters mitigates replay attacks by putting a current timestamp $T_s$ in the server request and integrating a server-generated random Nonce into the HMAC computation of the Merkle root. Replay attacks involve an adversary intercepting and retransmitting a previously valid message to impersonate a legitimate disrupt system operations. In order to compensate for this, the central server generates a fresh, random Nonce for each attestation session. This Nonce is concatenated with the Merkle root and processed using HMAC with a symmetric key, ensuring a secure and unique attestation signature for every session.

The process begins with the server generating an attestation request. The timestamp in this message prevents old requests to be replayed. In addition, the meter does not replay to a request with repetitive $T_s$. The server also sends a freshly generated Nonce to the smart meter. The smart meter concatenates the received Nonce with the computed Merkle root and applies HMAC using a pre-shared symmetric key. The resulting HMAC serves as a cryptographic proof that securely binds the sensitive data (represented by the Merkle root) to the session-specific Nonce.

The smart meter transmits the concatenated Merkle root and Nonce, along with the computed HMAC, back to the server. Upon receiving the response, the server verifies the HMAC by replicating the concatenation and HMAC computation using the Nonce it issued and the shared symmetric key. The attestation is considered valid if the calculated HMAC matches the one received. Replay attempts fail because the server-generated Nonce is unique for every session, and a previously valid message would produce an incorrect HMAC when verified with a new Nonce.

This approach ensures the security of the remote attestation process by effectively binding sensitive data to a server-controlled session-specific Nonce. By employing HMAC with a symmetric key and enforcing Nonce uniqueness, the system guarantees the integrity and authenticity of smart meter communications while robustly preventing replay attacks.

\subsubsection{Denial of Service (DOS)}
Our proposed method is immune to DoS attacks because only authorized entities can generate valid requests. The verification of each request is a lightweight process, and the requests are rate-limited. In particular, the server includes a timestamp and a keyed HMAC in its initial request for attestation. An attacker cannot manipulate server requests because the verification process is straightforward on the smart meter side. In addition, the meter does not respond to requests with repeated timestamps. Verifying smart meter responses is also lightweight on the server side because the server drops messages with old Nonces, and HMAC calculation is efficient. By the way, our proposal is agnostic to the lower layers of the network, and those layers should already be resilient to DoS attacks.

\subsubsection{Other Attacks}
Our method is resilient to other potential attacks, including eavesdropping (due to encrypted communication), physical tampering (protected by the TEE), and key management issues (ensured through secure pre-shared symmetric keys). The approach also safeguards against malware or firmware exploits (because of attestation and verification of updates), enhancing smart grid systems’ overall security and reliability. This comprehensive defense mechanism makes the system robust against various adversarial scenarios, ensuring trust and reliability in smart meter communications.

\subsection{The Smart Meter Lifecycle}

A smart meter passes several stages before it ever reports a reading, and each stage introduces its own trust boundary. Understanding this lifecycle clarifies where our scheme places its trust and where an adversary might try to intervene.

\subsubsection{Manufacturing and Hardware Assembly}

The metering module, the MCU, and the supporting board are assembled in this stage. For our approach to work, the hardware must support TrustZone---e.g., be equipped with an ARM Cortex-M processor---so that a hardware-isolated execution environment is available from the very beginning.

\subsubsection{Firmware and Cryptographic Key Provisioning}

Once the hardware is ready, the initial firmware is installed, and each meter is
assigned a unique device-specific symmetric attestation key during a trusted
manufacturing or commissioning process. The device-side key is provisioned into
the TrustZone-backed secure storage of the TEE rather than being stored in
general-purpose flash. The corresponding key record is securely registered with
the Distribution Management System (DMS) through the trusted provisioning
infrastructure.

During normal operation, the device-side key remains within the Secure World and
is not accessible to the Normal World or untrusted application code. All
key-dependent attestation operations, including HMAC generation, are performed
within the TEE. This hardware-enforced isolation mitigates key-extraction risks
arising from Normal-World compromise and ordinary software-based attacks, while
enabling subsequent attestation rounds without exposing the device secret.

\subsubsection{Device Identity and Factory Enrollment}

Before leaving the factory, each meter is assigned a unique device identifier and enrolled in the utility's asset-management system. This binding between identifier and provisioned key gives every unit a distinct, verifiable identity, which is what allows the operator to tell one meter apart from another and to reason about which devices are legitimately part of the network.

\subsubsection{Trusted Installation and Commissioning}

Deployment should be carried out by trusted installers following defined procedures, not treated as a purely mechanical task. During commissioning the meter is physically connected, activated, and onboarded to the network, and its identity is checked against the records created at enrollment. Restricting who can install and activate a device closes a gap that would otherwise let a tampered or counterfeit unit slip into the system unnoticed.

\subsubsection{Key Lifecycle Management: Rotation and Revocation}
\label{subsec:key-lifecycle}

A symmetric key is only as safe as the process that manages it over time. In our model the operator can rotate keys periodically and blacklist the key of any meter that is decommissioned or found to be compromised. When a meter leaves the network for either reason, it can no longer produce a valid proof. This keeps the integrity structure consistent with the current and trusted set of devices.

\subsubsection{Trust Assumptions and Attack Surface Across the Lifecycle}

Taken together, these stages define who is trusted and when. We assume the manufacturer and the provisioning process are trusted, and that the secret key is protected inside the TEE from the moment it is generated. The main residual exposure lies in the handling and logistics window before installation and in the correctness of the installation step itself, which is why device identity checks and trusted commissioning matter as much as the runtime protections. The remainder of the paper focuses on the operational stage, where attestation and the Merkle tree defend data integrity once the meter is live.

\section {Future Work}
Our main goal is to ensure the integrity of smart grids by using secure computing such as Trusted Execution Environment. Considering the resource constrained devices, in this paper we have shown the feasibility of remote attestation in such devices by strengthening the attestation techniques with Merkle tree. However, this constraint does not apply to all other smart grid devices, and in our future work we will present hybrid attestation techniques to ensure integrity in a variety of other smart grid devices.

\section{Conclusion}
\label{sec:conclusion}

In this paper, we evaluated a standard smart meter with its built-in security features and found that these were insufficient. 
Therefore, we searched for a trustworthy execution environment that could support routine tasks while enhancing security. 
This led us to implement a smart electricity meter in the OP-TEE secure environment that achieves a higher level of security than conventional smart electricity meters.
The OP-TEE environment is based on ARM TrustZone, which provides us with a secure world and a normal world for operation. Nevertheless, our goal was even higher security.
In this trusted execution environment, we used remote attestation between the smart meter and the server.

The server will verify this protocol to establish communication and ensure that any change to the hashed data will cause the remote attestation to fail. 
We also used a Merkle Tree to aggregate and hash data from smart meters. We tested the Merkle Tree with different leaf counts, mapped it to a smart meter, and requested the hashed data from the server.
The Merkle root sent to the server requires successful remote attestation before transmission.
These measures provided an acceptable level of security and prevented direct access to the smart meter data. 
Our proposed method is efficient and secure against various attacks, has an acceptable footprint, and was shown to be superior to other related works and, therefore, suitable for resource-constrained devices, including smart meters.

\section{Acknowledgment}
The authors used Grammarly and AI-assisted tools during the writing process of this manuscript to improve the linguistic quality, check grammar, and refine the technical prose. After utilizing these tools, the authors thoroughly reviewed, edited, and verified the final content and take full responsibility for the accuracy and integrity of the published work.

\printcredits

\bibliographystyle{cas-model2-names}

\bibliography{ref}

\pagebreak
\bio{FIG_BIB_SHAKIBA}
Amirhossein Shakiba is currently a Ph.D. student in Computer Engineering at Shahid Beheshti University. he received his Master of Science in computer engineering at Tarbiat Modares University in September 2024. He was a part of the Blockchain Laboratory at Tarbiat Modares University under the supervision of Dr. Sadegh Dorri Noghoorani.
\endbio

\bio{FIG_BIB_GHAEINI}
Hamid Reza Ghaeini is a faculty member at the University of Tehran. He earned his Ph.D. in Information Systems Engineering and Design from the Singapore University of Technology and Design in 2019. A professional member of both IEEE and ACM, he serves on program committees, editorial boards, and as a reviewer for numerous prestigious international conferences and journals. He is also actively engaged with industry as a consultant and entrepreneur.
\endbio

\bio{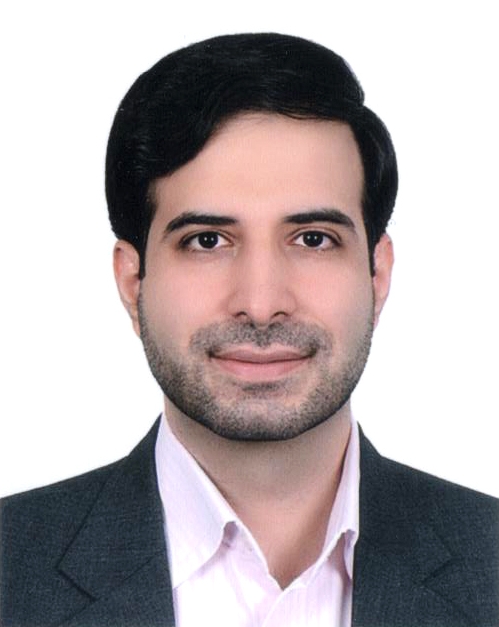}
Sadegh Dorri Nogoorani received his Ph.D. in Computer Engineering from Sharif University of Technology in 2016 and joined Tarbiat Modares University in 2017. His research interests include blockchain and distributed ledger technology, cybersecurity, and privacy in distributed systems.
\endbio

\end{document}